\documentclass{ametsocV6.1}

\usepackage{tabularx}
\usepackage{booktabs}
\usepackage{ragged2e}
\usepackage{hyperref}
\usepackage{xcolor}

\title{Transfer Learning for Dead Fuel Moisture Prediction Using Time-Warped Recurrent Neural Networks}

\authors{Jonathon Hirschi,\aff{a}\correspondingauthor{Jonathon Hirschi, jonathon.hirschi@ucdenver.edu}
Jan Mandel\aff{a} 
Adam Kochanski,\aff{b} 
}

\affiliation{\aff{a}{Department of Mathematical and Statistical Sciences, University of Colorado Denver, Colorado}\\
\aff{b}{Department of Meteorology and Climate Science, Wildfire Interdisciplinary Research Center, San Jose State University, San Jose, California}\\
}

\abstract{
This paper proposes a time-warping transfer learning method, a technique for temporally rescaling the learned dynamics of a recurrent neural network (RNN) with a Long Short-Term Memory (LSTM) layer to enable task transfer across fuel moisture classes. Fuel moisture content (FMC) is divided into idealized classes based on characteristic lag time. Large quantities of real-time data are available for 10h fuels from sensors on weather stations, but observations of other fuel classes are sparse in space and time. We use transfer learning to adapt an RNN pretrained on 10h FMC to predict FMC for 1h, 100h, and 1000h fuels. We validate this method using data from a landmark field study conducted in Oklahoma that was used to calibrate the state-of-the-art Nelson fuel moisture model.
} 

\begin{document}

\maketitle

%
%
%
\statement
Dead fuel moisture data are widely available for 10-hour fuels but scarce for 1, 100, and 1000-hour fuel classes, making standard machine learning approaches difficult to train reliably for the sparsely observed classes. We develop a transfer learning method based on time-warping to adapt an RNN trained on 10-hour fuel data to predict 1-, 100-, and 1000-hour fuel moisture, where target data are scarce, while modifying only a small subset of model parameters. To our knowledge, this is the first application of transfer learning to fuel moisture prediction and the first published ML study to target 100-hour and 1000-hour fuel moisture. It shows that transfer learning can extend data-driven prediction beyond the fuel class with abundant sensor observations.

%
%

%

\section{Introduction}

Fuel moisture content (FMC) is a measure of the water content of fuels that is important for understanding wildfire danger and active fire behavior, e.g.~\citep{NCEI-2024-DFM}.  It is expressed as the ratio of the mass of water in the fuel to the dry mass of the fuel, typically given as a percentage. 

Live and dead fuels are typically modeled separately because they are governed by different processes. Dead fuel moisture content is controlled primarily by external environmental conditions such as temperature, relative humidity, solar radiation, and wind, through moisture exchange with the atmosphere (absorption and desorption). In contrast, live fuel moisture content is regulated by internal physiological processes within the plant, including water uptake from soil, storage, and transpiration~\citep{NCEI-2024-DFM, Lewis-2024-DFM}. This paper focuses on predicting dead FMC as a response to weather.

Operational wildfire management tools and wildfire research commonly use the idealized 1-hour, 10-hour, 100-hour, and 1000-hour fuel classes—denoted as FM1, FM10, FM100, and FM1000—which correspond to characteristic dead fuel size classes of approximately 0–0.25 inches (0–0.6 cm), 0.25–1 inch (0.6–2.5 cm), 1–3 inches (2.5–7.6 cm), and 3–8 inches (7.6–20 cm) in diameter, respectively. Fuel classes are defined by their time lag, the characteristic time required for a fuel’s moisture content to change by $1-e^{-1}$ (about 63\%) of the difference between its initial moisture content and its new equilibrium value in response to a change in atmospheric conditions~\citep{Byram-1963-ADP,Finney-1998-FFA}. Large quantities of data exist for fuels with a lag time of 10 hours, but observations for other lag times are sparse.

A large network of Remote Automatic Weather Stations (RAWS) provides data from standardized sensors for FM10~\citep{NIFC-2024-RAW}. The FMC sensor consists of a \mbox{1/2-inch} pine dowel with a probe that estimates the moisture content in the woody interior~\citep{Campbell-2015-CFM, FTS-2016-FSS}. RAWS across the continental United States (CONUS) provide real-time data for FM10~\citep{Synoptic-2025-MSN}.  

Observations of the other fuel classes are sparse in space and time, as automated sensors are only available for FM10. For all fuel classes, FMC can be measured gravimetrically by weighing fuel samples before and after oven drying. Gravimetric measurement is accurate and direct, but it is slow and labor-intensive, so the available data is much sparser. These observations include field samples collected by wildfire management agencies and distributed through the Fire Environment Mapping System (FEMS)~\citep{USDA-FEMS-2024}, which can be accessed through the National Fuel Moisture Database (NFMDB)~\citep{WIRC-2026-WFMP}.

Controlled field studies provide higher-temporal-resolution observations, with fuel dimensions standardized to the conventional time lag fuel classes. A particularly important controlled field study is that of \citet{Carlson-2007-ANM}, which measured FMC for multiple fuel classes up to twice daily from spring 1996 through 1997. From now on we refer to that study as the Oklahoma field study. Those data were used to calibrate the \citet{Nelson-2000-PDC} fuel moisture model for use within the National Fire Danger Rating System (NFDRS), as well as other process-based models~\citep{vanderKamp-2017-MFS}, and we use them here because they are historically important and have higher temporal resolution than the FEMS field studies.

Modern machine learning (ML) models use large amounts of FM10 data to learn spatial and temporal patterns and provide accurate forecasts. In contrast, ML approaches to predicting fine fuel moisture (FM1) and FMC for larger fuels can rely only on the use of relatively small amounts of data. Traditional physics-based models of FMC continue to be the standard approach to predicting FM1, FM100, and FM1000. This paper attempts to fill a gap in the FMC modeling literature by leveraging the vast network of FM10 sensors to improve predictions for other fuel classes. We describe a new approach to transfer learning based on modifying the learned temporal dynamics of a recurrent neural network (RNN), which we refer to as time-warping. We use a pretrained RNN that has been developed to predict FM10, then apply the time-warping method to predict FM1, FM100, and FM1000. We compare this method to several standard transfer learning methods and show that we achieve similar levels of accuracy despite modifying only a small fraction of the learned parameters relative to the other methods.

This paper is based on the thesis~\citet{Hirschi-2026-TWR}, which builds on forecasting of FMC by RNNs with weather inputs, introduced by~\citet{Mandel-2023-BFM} and developed into a spatiotemporal product by~\cite{Hirschi-2026-RNN}.
 
\subsection{Modeling Approaches for Dead Fuel Moisture Content}

Mathematical models of FMC can generally be divided into three categories: physical process-based models, time lag models, and ML models. Process-based models and modern ML approaches still use fuel classes to define output variables, as that remains the standard for data collection~\citep{Zahn-2011-SFM}. In some areas of the literature, researchers study the moisture content of fine fuels, which includes dead bark, twigs, and leaf litter. FM1 has a specific time lag definition and has been studied with standardized wooden dowels, while fine fuel moisture content is used more generally in wildfire science.

\subsubsection{Physical Process-Based Models}

Physical process-based methods attempt to directly represent the transfer of energy and moisture between a wood stick and its environment. 
\citet{Nelson-2000-PDC} developed a model for FMC as part of the NFDRS that solves coupled one-dimensional heat and moisture transport equations in the radial coordinate of a cylindrical wood stick. The model receives weather inputs of air temperature, RH, accumulated rainfall, and downward shortwave solar radiation. 
\citet{vanderKamp-2017-MFS} modeled FMC using two homogeneous layers and added the effect of the wind to some of the formulations in the Nelson model. Both models were calibrated using the Oklahoma field study.
\citet{Matthews-2006-PBM} proposed a process-based model for fine fuel moisture based on coupled one-dimensional energy and water balance equations through a litter layer, with boundary conditions at the atmosphere above and the soil below. The model was calibrated using fine fuel moisture samples from two locations in Australia.

The Nelson model is currently used within the NFDRS to predict FMC for multiple dead fuel classes~\citep{Jolly-2024-MUN}. The same mathematical model is used for different fuel classes, with the assumed stick radius changed to match the dimensions of the idealized fuels. The Nelson model is considered by many researchers to be the state-of-the-art method and is widely adopted in operational contexts. 

\subsubsection{Time Lag Models}

Time lag models for predicting FMC have been utilized for decades by researchers and in operational use by agencies. The concept of time lag originates in the diffusion-based drying analysis of forest fuels~\citep{Byram-1963-ADP}. Time lag models treat a fuel element as homogeneous and assume that the FMC approaches a dynamically-changing equilibrium moisture content. Across fuel classes, the same first-order relaxation model is used with different characteristic time lags~\citep{Finney-1998-FFA}.

\citet{Fosberg-1971-DHT} developed operational time lag equations for predicting the daily change in FM1 and FM10 using equilibrium moisture content calculated from mid-day temperature and RH. The Canadian Forest Fire Weather Index system developed the Fine Fuel Moisture Code, which models the daily change in the FMC for fine fuels using time lag equations with the wetting and drying equilibria~\citep{VanWagner-1987-DSC}.

\citep{Mandel-2014-RAA} developed a system of time lag ordinary differential equations (ODEs) to forecast FMC. Data assimilation of FM10 observations from RAWS is transferred to FM1 and FM100~\citet{Vejmelka-2014-DAF,Vejmelka-2016-DAD}. This model is currently used operationally within WRF-SFIRE, a coupled atmosphere-wildfire model. The present paper is part of a research effort to improve the FMC forecasting within WRF-SFIRE.

\subsubsection{Machine Learning Models}

There has been growing interest in using ML models built directly from data. We use the term \emph{static} models to refer to any non-recurrent ML model, where models learn the instantaneous relationship between the weather and FMC, without using information about the past FMC of the fuel. 

To predict FM10, \citet{McCandless-2020-EWS} developed a gridded daily product for 10-h dead fuel moisture content using static ML models. In a followup study, \citet{Schreck-2023-MLV} developed a static XGBoost model using satellite and numerical weather prediction (NWP) inputs, producing a data product used for reanalysis and wildfire simulation initialization. Other studies have applied static random forest models with weather inputs from ground-based stations~\citep{Lee-2020-EFM, Chae-2025-MLB}. Recurrent models have also been explored; \citet{Fan-2021-PGD} used an LSTM to predict FM10 using outputs from \citet{vanderKamp-2017-MFS} as inputs. In prior work, we developed an LSTM using NWP inputs and evaluated its performance in a large-scale spatiotemporal forecasting analysis~\citep{Hirschi-2026-RNN}, showing improved accuracy relative to baseline methods; that model provides the starting point for the transfer-learning analysis in this paper.

Fine fuel moisture content is often represented operationally as FM1~\citep{Jolly-2024-MUN}. Static models, including linear regression and random forests, have been applied in forest environments in southern China~\citep{Hou-2024-CAM} and across a range of ecosystems in Australia~\citep{Nolan-2016-PDF}. \citet{Fan-2023-CFM} trained static models using samples from forests in northern China. Recent applications of RNNs include \citet{Han-2026-KEM}, who propose a physics-constrained RNN, showing improved performance under data-limited conditions, while standard LSTMs perform best when trained on the full dataset.

\subsection{Transfer Learning in Environmental Science}

Transfer learning is a general term in ML for reusing knowledge learned in one context in another. The main practical motivation is when one modeling problem has insufficient observed data to adequately train an ML model, but more data are available for a related problem. It can also help a model learn deeper aspects of a system that may generalize across related problems.

\citet{Pan-2010-STL} formulated transfer learning, and their terminology is widely used today. The \emph{domain} of a problem consists of the inputs available to the model. The \emph{task} describes what the model predicts. In transfer learning, a model is trained first on a \emph{source} domain and a task, then adapted to a \emph{target} domain and task. Different forms of transfer learning differ in what is carried over from the source model and what is allowed to change when the model adapts to the target.

This paper focuses on parameter-based transfer learning for neural networks. We use \emph{pretrained} to describe a model fit on the source domain and task, and \emph{fine-tuning} to describe updating its parameters using target data. The phrase \emph{zero-shot} refers to the case where no fine-tuning is performed, and predictions for the target task are generated without any training on target data. Freezing layers in neural networks is a method of fine-tuning that preserves parameters that are expected to be shared across learning tasks, e.g.,~\citet{Yosinski-HTF-2014}. Freezing a layer fixes its parameter values after pretraining, so they are not updated during fine-tuning. 

Transfer learning has been extensively studied in image processing with convolutional neural networks (CNNs), often by freezing early layers and fine-tuning downstream layers~\citep{Yosinski-HTF-2014}. In environmental science, these methods have become increasingly common for image-based problems~\citep{Ma-2024-TLE}, especially when models are trained in regions with abundant data and then adapted to regions with sparse data. They have also been used to account for differences among sensing platforms, such as different satellite networks or aerial imaging systems. Within wildfire science, transfer learning has been used primarily for image-based wildfire detection by pretraining CNNs on general image datasets and fine-tuning them on wildfire imagery~\citep{Sousa-2020-WDT, Hong-2024-WDV}.

Transfer learning with RNNs has received comparatively less attention, but several studies in environmental science are relevant to the methods used in this paper.
In streamflow modeling, transfer learning with LSTMs has been used to train models in data-rich regions and transfer them to data-sparsely observed regions~\citep{Ma-2021-THD, Khoshkalam-2023-ATL}. Existing work has primarily focused on transferring models between locations for the same prediction task, with more limited work on transferring models between different output variables.

\subsection{Time-Warping ML Sequence Models}

In classical ordinary differential equations (ODEs), temporal rescaling changes the units of time in the model equations through a transformation $t \mapsto \gamma t$ for $\gamma>0$. For linear first-order ODEs, the same transformation changes the rate of change of the system, corresponding to faster or slower dynamics. We refer to this as time-warping to distinguish it from standard temporal rescaling, where the underlying output variable is unchanged.

Time-warping is a type of transfer learning, in which related problems share the same inputs but differ mainly in the rate at which their dynamics evolve. Under the assumption that the rate of change of FMC is the only difference between fuel classes, the learning problems share the same input variables while the tasks differ by time scale. In the time lag models of FMC, the system is time-warped to produce predictions of the different fuel classes. 

Time-warping in machine learning has primarily been used as a preprocessing or alignment technique. In contrast, there has not been extensive research into post-hoc manipulation or time-warping of learned dynamics, which is the focus of this paper. 

The Chrono-Initialization method for LSTMs initializes bias parameters based on an assumed range of memory time constants~\citep{Tallec-2018-CRN}. It identifies the forget and input gates as controlling the effective time scale of the learned dependencies. \citet{Ohno-2021-RNN} revisit this interpretation in a more general setting and argue that the forget gate represents the time scale of the state.

\citet{Quax-2020-ATS} add learnable parameters to an RNN that allow each unit to adapt the rate at which the hidden state evolves. The authors interpret this as learning the intrinsic time scales of continuous-time dynamic systems. Empirically, they show that the architecture learns time constants that reflect the temporal structure of different learning tasks. 

Dynamic time-warping (DTW) is a method for comparing the similarity of discrete time series, originally developed for speech recognition~\citep{Sakoe-1978-DPA}. Despite the similar terminology, it is not time-warping as defined in this paper, because it does not involve a global temporal rescaling. DTW aligns two time series by applying a loss function to pairs of time indices to form a loss matrix and computing the minimum-cost path through that matrix. Recent studies have used DTW to select representative source datasets before LSTM fine-tuning~\citep{Duan-2026-DTW} and to define evaluation metrics for LSTM forecasts of geomagnetic properties~\citep{Laperre-2020-DTW}.

\subsection{Research Objectives}

In this paper, we predict FMC for different fuel classes using a novel transfer learning method. While simple time lag ODEs can have less predictive accuracy than RNNs, we use principles from the time lag system to develop a method for modifying the learned temporal dynamics of a pretrained RNN with an LSTM recurrent layer. The RNN is used here as a learned dynamical system: it maps a time series of weather inputs to a time series of fuel-moisture predictions through a recurrent internal state carried forward in time. Starting with a pretrained RNN for predicting FM10, we modify a set of parameters in the LSTM layer to generate predictions for FM1, FM100, and FM1000. The model predictions are compared to the observations from the Oklahoma field study. We compare the accuracy from the time-warping method to established baseline methods of transfer learning, and additionally we compare to the reported accuracy metrics for the Nelson model from the Oklahoma field study which was used to calibrate the model~\citep{Carlson-2007-ANM}.

\section{Methods}

\begin{figure}[tbph]
\centering
\includegraphics[width=12cm]{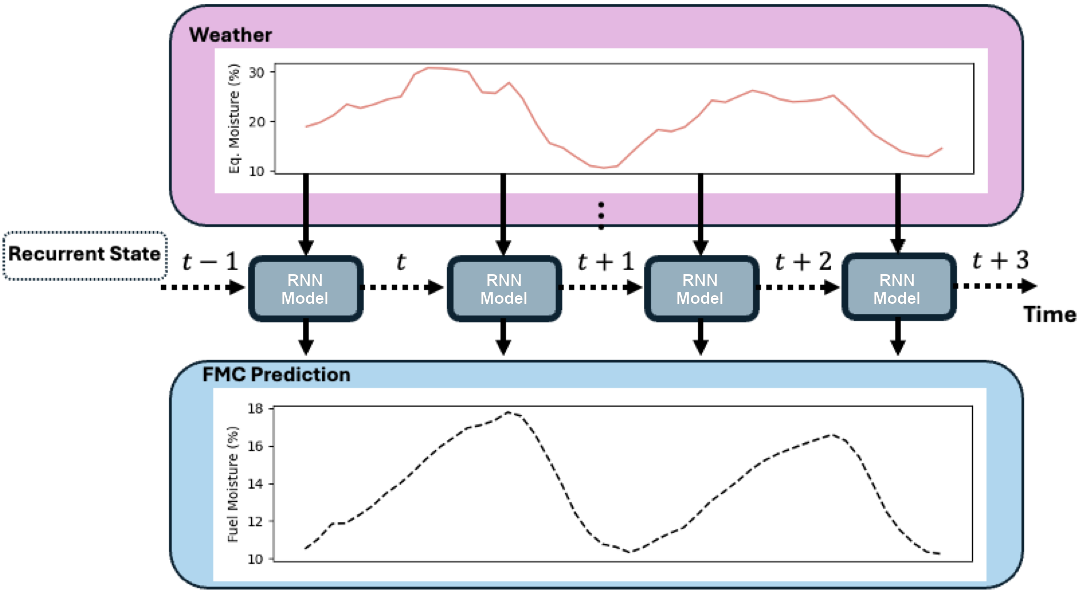}
\par\vspace{0.8cm}
\includegraphics[width=12cm]{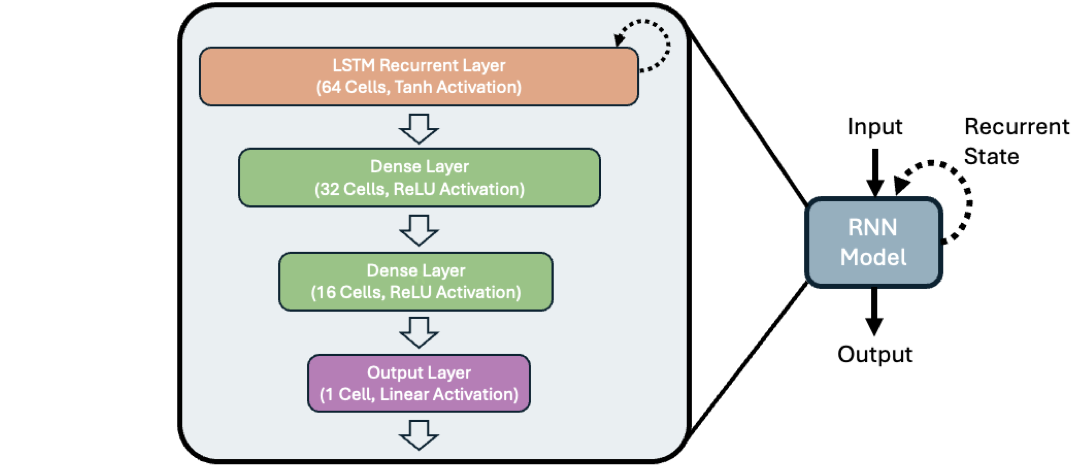}

\caption[RNN Architecture and Computational Flow.]{The RNN model of FMC. Top:  The processing unrolled in time. The RNN model is applied at each time step to map a~time series of weather inputs to a~time series of FMC at the same time steps, with the recurrent state propagated forward.
Bottom: Detail of the RNN model architecture with one LSTM layer and three dense layers. 
Adapted from~\citet{Hirschi-2026-RNN}.
}
\label{fig:rnn_model}
\end{figure}

We used the RNN model from~\citet{Hirschi-2026-RNN} (Figure~\ref{fig:rnn_model}), trained on all available FM10 observations from RAWS within the Rocky Mountain region from 2023 to 2024, using truncated backpropagation through time \citep[e.g.,][]{Zhang-2023-DDL}. The validation set consists of the time series in the last 48-hour period in 2024 in a random sample of $10\%$ of locations, and it is used only to control the early stopping procedure. To characterize variability in model training, we generated 100 realizations of the RNN training procedure by randomly varying the validation set, the order of the training samples, and the initial weights. Each realization starts from a different random initialization and produces a different set of trained model parameters for the same architecture.

\subsection{Study Area and Data}

The Oklahoma field study was conducted near the town of Slapout, Oklahoma, from spring 1996 through 1997.
Arrays of pine dowels representing the different fuel classes were measured up to twice daily, usually around 13:00 and 22:00 \citep{Carlson-2007-ANM}.
Weather observations from a nearby ground-based station in the Oklahoma Mesonet~\citep{McPherson-2007-OMU} provided the model inputs~\citep{OKMeso-2026-MSI}.
Table~\ref{tab:all_vars} shows summary statistics for the hourly weather from the Mesonet station over the same time period as the observed FMC.

\begin{table}[htbp]
\caption[Oklahoma Field Study Weather Data Summary.]{Oklahoma Field Study Weather Data Summary. Weather variables and geographic features are from the Slapout Mesonet station. The time period is from March 26, 1996 through the end of 1997.}
\label{tab:all_vars}
\small
\begin{tabularx}{\textwidth}{l l p{4cm} p{1.5cm} p{2.5cm}}
\toprule
\textbf{Variable} & \textbf{Units} & \textbf{Description} & \textbf{Mean} & \textbf{(Low, High)} \\
\midrule
Drying Equilibrium & $\%$ & Derived from RH and temperature. & 18.02 & (1.29, 60.56) \\
\midrule
Wetting Equilibrium & $\%$ & Derived from RH and temperature. & 16.53 & (0.61, 56.10) \\
\midrule
Solar Radiation & $\text{W\,m}^{-2}$ & Downward shortwave radiative flux. & 208.41 & (0, 1177) \\
\midrule
Wind Speed & $\text{m\,s}^{-1}$ & Wind speed at 10m. & 2.52 & (0.40, 8.61) \\
\midrule
Rain & $\text{mm\,h}^{-1}$ & Rain accumulation over the hour. & 0.08 & (0.00, 42.17) \\
\midrule
Hour of Day & hours & Hour of day, UTC time. & 11.50 & (0, 23) \\
\midrule
Day of Year & days & Day of year, UTC time. & 201.70 & (1, 366) \\
\midrule
Elevation & meters & Height above sea level. & 774.00 & - \\
\midrule
Longitude & degree & X-coordinate & -100.26 & - \\
\midrule
Latitude & degree & Y-coordinate & 36.60 & - \\
\bottomrule
\end{tabularx}
\end{table}

Geographic predictors include the elevation, longitude, and latitude from the station location. The hour of the day and the day of the year are used as temporal predictors.

Table~\ref{tab:fmc_summary} provides a summary of the sample sizes and typical FMC values for the different fuel classes.
 As the characteristic time lag increases for the fuel classes, the typical FMC values decrease.
 Larger fuels tend to have lower average moisture content. 

\begin{table}[htbp]
\caption[Oklahoma Field Study FMC Data Summary]{Oklahoma Field Study FMC Data Summary. Observations are made with gravimetric measurements on arrays of wooden dowels. The time period is from March 26, 1996, through the end of 1997.}
\label{tab:fmc_summary}
\centering
\begin{tabular}{lccc}
\hline
Fuel Class & N. Observations & Mean & (Low, High) \\
\hline
FM1    & 1,233 & 15.33 & (0.0, 109.2) \\
FM10   & 1,232 & 15.02 & (1.6, 64.3) \\
FM100  & 871   & 13.44 & (5.4, 35.7) \\
FM1000 & 874 & 11.37 & (4.7, 27.9) \\
\hline
\end{tabular}
\end{table}

\subsection{Zero-Shot Transfer Learning for FM10}

Before evaluating transfer learning methods that change the target learning task, we consider a case where the source and target are both intended to predict FM10, but the input and observation data sources differ. The RNN is trained on the source data using weather inputs from NWP and FM10 observations from RAWS sensors, then evaluated on the target data using weather inputs from ground-based stations and gravimetric FMC measurements from the Oklahoma field study. The different data sources induce different error distributions. FMC predictions are generated for each of the 100 RNN realizations. These are zero-shot predictions, since no fine-tuning is done to adapt the RNN to the target data. The zero-shot predictions are evaluated against the observed FM10 values, and the model accuracy is compared to the Nelson model in Section~\ref{sec:zeroshot}.

\subsection{The Time-Warping Method of Transfer Learning}

We propose a new method of transfer learning that attempts to modify the learned temporal dynamics of a pretrained LSTM.
 The time-warping method of transfer learning is inspired by the standard temporal rescaling in a time lag model.

\subsubsection{Time-Warping in Time Lag Systems}

Consider a simple time lag model of FMC with continuous-time state $m(t)$ at time $t$ and one external input, the equilibrium moisture content $X(t)$~\citep{Viney-1991-RFF},
\begin{equation}
\label{eq:tcont_fmc}
\frac{d}{dt}m(t) = \frac{X(t)-m(t)}{\tau}. 
\end{equation}
Here, $\tau$ denotes the characteristic time lag in hours for a fuel class.
 For the standard fuel classes, $\tau$ is associated with the nominal lag times 1, 10, 100, and 1000 hours.
 
 If we use a time step of 1 hour, assume that the equilibrium moisture content is constant within each interval $(t-1,t]$, $t=1,2,\ldots $, and define the discrete-time variables $m_t = m(t)$ and $X_t = X(t)$ for $t=0,1,2,\ldots$, then \eqref{eq:tcont_fmc} has the solution
\begin{equation}\label{eq:tl_fmc}
\begin{aligned}
    m_{t} &= a\, m_{t-1} + (1-a) X_t, 
    \qquad a=e^{-1/\tau}, \qquad t=0,1,2,\ldots .   
\end{aligned}
\end{equation}
By selecting an initial FMC, denoted by $m_0$, the recursion~\eqref{eq:tl_fmc} generates a unique sequence of values $m_t, \ t=1,2,\ldots$.
 The coefficient $a=e^{-1/\tau}$ lies in the interval $(0,1)$, so~\eqref{eq:tl_fmc} is a weighted average between the previous state of the system and the current input. 

We define a time-warp by a factor $\gamma >0$ through the transformation $t \mapsto \gamma t$.
 This changes the time scale of the original system.
 Denote the time-warped solution by $\widetilde m(t)=m(\gamma t)$.
 In the discrete-time system \eqref{eq:tl_fmc}, this transformation raises the parameter $a$ to the power $\gamma$,
\begin{equation}
\label{eq:timewarp}
\begin{aligned}
    \widetilde m_{t} = a^\gamma \widetilde m_{t-1} + (1-a^\gamma)X_t, \qquad \gamma>0, \;a=e^{-1/\tau}.
\end{aligned}
\end{equation}

\subsubsection{Time-Warping an LSTM}

\begin{figure}[tbh]
\centering
\includegraphics[width=12 cm]{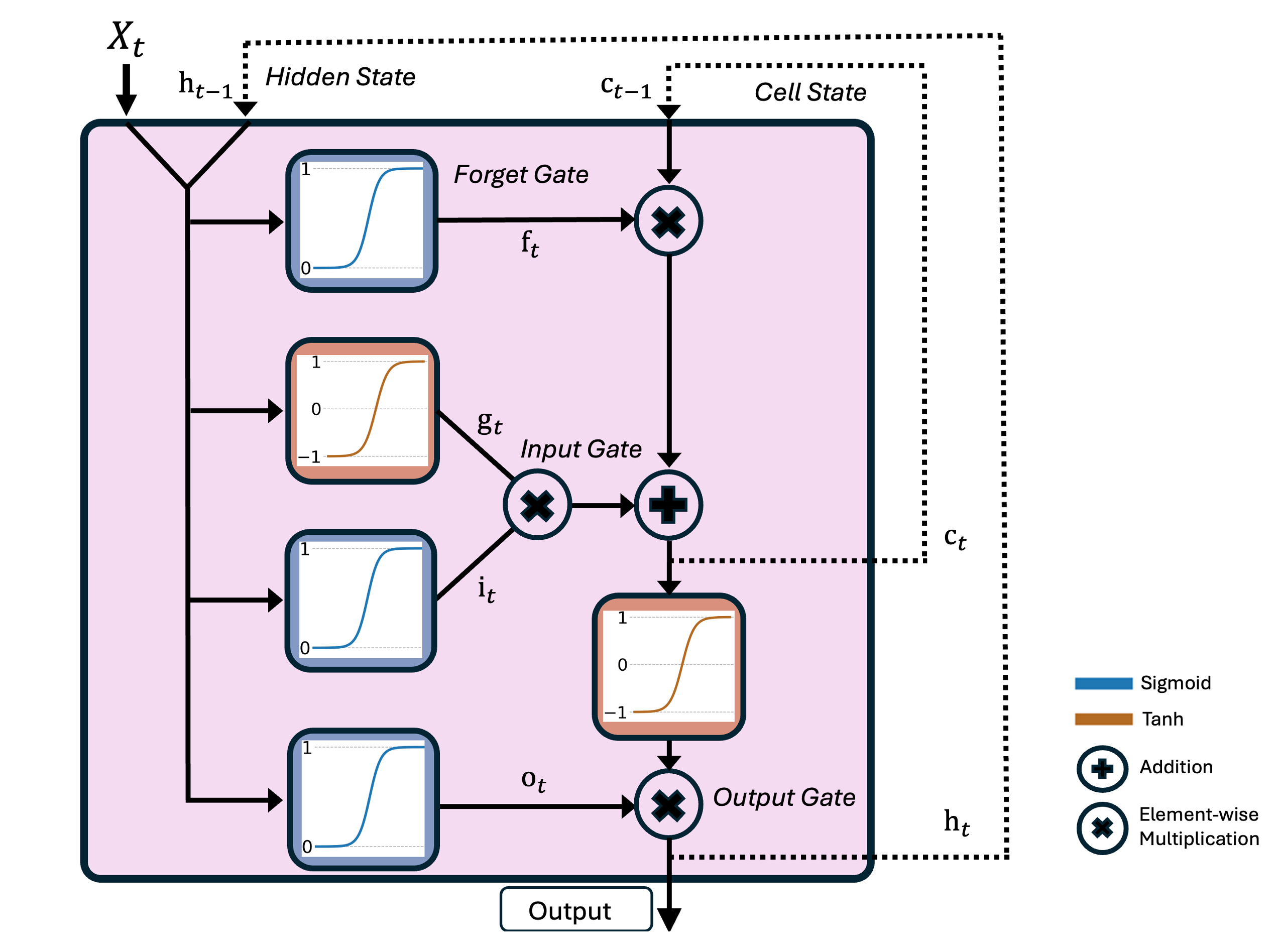}
\vspace{0.75em}
\begin{equation*}
\begin{aligned}
f_t &= \sigma(W_{xf}\,X_t + W_{hf}\,h_{t-1} + b_f) \\
i_t &= \sigma(W_{xi}\,X_t + W_{hi}\,h_{t-1} + b_i) \\
g_t &= \tanh(W_{xg}\,X_t + W_{hg}\,h_{t-1} + b_g) \\
o_t &= \sigma(W_{xo}\,X_t + W_{ho}\,h_{t-1} + b_o) \\
c_t &= f_t \otimes c_{t-1} + i_t \otimes g_t \\
h_t &= o_t\otimes \tanh(c_t)
\end{aligned}
\end{equation*}
\caption[A single LSTM unit.]{A single LSTM unit, following~\citet[Ch.~15]{Geron-2019-HOM}. Image adapted from~\citet{Hirschi-2026-RNN}.
} 
\label{fig:lstm_cell}
\end{figure}

The LSTM (Fig.\ref{fig:lstm_cell})  was developed to address computational issues that arise when training RNNs, known as the exploding and vanishing gradient problem~\citep{Hochreiter-1997-LST}. 
It maintains two states over time.
 The \emph{cell state}, denoted by $c_t$, acts as long-term memory, while the \emph{hidden state}, denoted by $h_t$, acts as short-term memory and is also the output of the LSTM. The activations of the forget, input and output gate are denoted by $f_t$, $i_t$, and $o_t$, respectively. 
 The products are standard matrix multiplications and $\otimes$ is element-wise multiplication.
 The forget gate controls how much of the previous cell state is retained, while the input gate controls how much new information is added.

For a time lag system~\eqref{eq:tl_fmc}, there exists a specially constructed single-unit LSTM with linear activation that approximates the system to arbitrary accuracy.
In that construction, the forget and input gate weights are zero and the activations are constant, $f_t=b_f=a$ and $i_t=b_i = 1-a$, thus determined by the bias terms $b_f$ and $b_i$ only.
By modifying $b_f$ and $b_i$, the corresponding time-warped system~\eqref{eq:timewarp} can be approximated with the same accuracy~\citep[pp.~52--56]{Hirschi-2026-TWR}.
This focus on the forget and input gate biases is also consistent with prior work linking these gates, or related recurrent timing parameters, to characteristic time scales in recurrent models~\citep{Tallec-2018-CRN,Ohno-2021-RNN,Quax-2020-ATS}.

This result is used to motivate the time-warping method.
We start with an RNN that is pretrained on a source learning task with one characteristic time scale, resulting in bias parameters $b_f$ and $b_i$.
For a time-warp factor $\gamma>0$, let $\alpha_f(\gamma)$ and $\alpha_i(\gamma)$ denote additive shifts of the forget and input gate bias parameters,
\begin{equation}
\label{eq:timewarped_bias}
\begin{aligned}
    \widetilde b_f &= b_f + \alpha_f(\gamma) \\
    \widetilde b_i &= b_i + \alpha_i(\gamma),
\end{aligned}
\end{equation}
where shifting the bias terms increases or decreases the associated gate activation.
This changes the dependence on the past state of the system versus the dependence on the current input data.
In the simple single-unit construction above, slowing down the dynamics is associated with increasing the forget-gate activation and decreasing the input-gate activation, while speeding up the dynamics is associated with the opposite pattern.
In that construction, the required bias modifications are known functions of $\gamma$.
For an arbitrary pretrained LSTM, however, the assumptions of the mathematical proof do not hold, and the exact time-warping factor is not known.
Accordingly, we use the bias-shifting mechanism only as a~motivation and treat $\alpha_f$ and $\alpha_i$ as two scalar tuning parameters, applied uniformly to all units in the LSTM layer and at all times, and fit empirically from target-task data.
More flexible choices would also be possible.
Table~\ref{tab:twarp_algorithm} describes an algorithm in which grid search is used to fit the two bias shifting parameters to data from the target learning task. 

\begin{table}[htbp]
\centering
\caption[Algorithm for time-warping a pretrained LSTM]{Algorithm for time-warping a pretrained LSTM, with bias-shift parameters selected by grid search.}
\label{tab:twarp_algorithm}
\begin{tabular}{p{0.95\linewidth}}
\hline
\textbf{Setup:} Pretrained LSTM with bias vectors $(b_f, b_i)$; target-data training set $\mathcal{D}_{\text{train}}$; test set $\mathcal{D}_{\text{test}}$; candidate bias shifts $(\alpha_f, \alpha_i)$ \\
\\
\textbf{1.} Use grid search on $\mathcal{D}_{\text{train}}$ to select bias shifts $(\alpha_f, \alpha_i)$ for the forget and input gates \\
\textbf{2.} Shift the LSTM biases: $\widetilde{b}_f = b_f + \alpha_f$, $\widetilde{b}_i = b_i + \alpha_i$ \\
\textbf{3.} Construct the time-warped model using the shifted biases \\
\textbf{4.} Generate predictions and evaluate performance on $\mathcal{D}_{\text{test}}$ \\
\\
\textbf{Output:} Test performance and selected bias shifts $(\alpha_f, \alpha_i)$ \\
\\
\textbf{Note:} If fine-tuning is used, it is applied after Step 3 using $\mathcal{D}_{\text{train}}$ before final evaluation on $\mathcal{D}_{\text{test}}$. \\
\hline
\end{tabular}
\end{table}

In this study, the time-warping algorithm used a grid search with 25 candidate values for each bias shifting parameter over the interval $(-5,5)$. The search was performed separately for FM1, FM100, and FM1000, yielding 625 candidate parameter combinations for each fuel class.

\subsection{Comparison of Transfer Learning Methods for FMC Prediction}

In our FMC application of transfer learning, the source learning task is predicting FM10, and the target learning tasks are predicting FM1, FM100, and FM1000. Their predictors are the same (Table~\ref{tab:all_vars}). However, the observed FMC data are much sparser for the target learning tasks. Although the source task utilized hourly observations of FM10 from a large network of physical locations, for the target learning task, the FMC was measured at most twice daily.

Two transfer learning methods that utilize the time-warping method, with and without fine-tuning, are evaluated. 
 The method with fine-tuning modifies the algorithm in Table~\ref{tab:twarp_algorithm} by using a validation set and fine-tuning all of the network weights after step 3.
 These are compared to established transfer learning techniques.
 All transfer learning scenarios utilize the same RNN model architecture shown in Figure~\ref{fig:rnn_model}.

Table~\ref{tab:tl_list} shows a summary of the transfer learning methods used in this analysis.
\emph{No transfer} describes directly training the RNN on the Oklahoma field data, with no pretraining.
 Then, \emph{full fine-tuning} describes taking the pretrained RNN and fine-tuning on the Oklahoma field data, allowing all model parameters to be updated.
 Two scenarios in which the layers are frozen are also used.
 \emph{Freezing the recurrent layer} allows only the parameters in the dense and output layers to be updated.
 The assumption of this method is that the early layers of a network extract a general set of latent features, and the downstream layers match the latent features with a particular response value.
 \emph{Freezing the dense layers} allows the parameters of the LSTM recurrent layer to be updated. Although this method is used as a baseline for comparison for the time-warp technique, it shares a motivation with the time-warping technique.
 Fine-tuning the parameters of the recurrent layer is meant to accomplish the same thing as the time-warping, but in a purely data-driven way rather than through direct modification of the input and forget gates that is motivated by theoretical considerations.
 
\begin{table}[htbp]
\centering
\caption[Transfer Learning Methods Comparison for FMC Prediction]{Transfer learning methods comparison for FMC prediction.}
\label{tab:tl_list}
\begin{tabularx}{\textwidth}{|>{\RaggedRight\arraybackslash}p{3.5cm}|>{\RaggedRight\arraybackslash}X|>{\RaggedRight\arraybackslash}X|}
\hline
\textbf{Method Name} & \textbf{Details} & \textbf{Literature Status} \\
\hline
No Transfer
& Random initialization and direct training on the target learning task.
& Standard supervised training baseline. \\
\hline
Full Fine-Tuning
& Model pretrained on a source task and all parameters updated on the target task.
& Standard transfer learning baseline. \\
\hline
Freeze Recurrent Layer
& Recurrent weights fixed while dense/output layers are retrained on the target task.
& Common transfer learning strategy, widely used for CNNs. \\
\hline
Freeze Dense/Output Layers
& Dense/output layers fixed while input and recurrent layers are retrained.
& Reverse-freezing baseline; rarely explored in literature. \\
\hline
Time-Warping
& Parameters for shifting the forget and input gate biases are fit, without additional training.
& New. \\
\hline
Time-Warping + Fine-Tuning
& Time-warp initialization followed by full fine-tuning on the target task.
& New. \\
\hline
\end{tabularx}
\end{table}

 The baseline methods that utilize fine-tuning use the same training procedure, with different initialization of parameters and different sets of parameters that are allowed to be updated during training.
 The accuracy metrics are $R^2$, the bias, and the RMSE.

\subsection{Experimental Design}

The Oklahoma field data consist of a single time series for each fuel class from one spatial location, so the data were split only in time. The training set was defined as one full year of data from the earliest observation so that, in principle, the models could learn seasonal variation. The remaining period was divided evenly into validation and test sets, and final accuracy metrics were calculated only on the test set. Tables~\ref{tab:data_split_summary} and~\ref{tab:data_split_counts} summarize the time periods and observation counts for the data splits.

\begin{table}[htbp]
\caption[Temporal data splits for Oklahoma field data.]{Temporal definition of the training, validation, and test splits for the Oklahoma field data. All dates are reported in UTC.}
\label{tab:data_split_summary}
\centering
\begin{tabular}{lccc}
\hline
 & Training Set & Validation Set & Test Set \\
\hline
Start Date (UTC) & 1996-03-26, 23:00 & 1997-03-27, 00:00 & 1997-08-13, 12:00 \\
End Date (UTC)   & 1997-03-26, 23:00 & 1997-08-13, 11:00 & 1997-12-30, 22:00 \\
\hline
\end{tabular}
\end{table}

\begin{table}[htbp]
\caption[Observation counts by temporal split for Oklahoma field data.]{Number of observations in the training, validation, and test splits for each variable in the Oklahoma field data. Note that not all analyses utilize the validation set.}
\label{tab:data_split_counts}
\centering
\begin{tabular}{lccc}
\hline
N. Observations & Training Set & Validation Set & Test Set \\
\hline
Weather & 8,761 & 3,348 & 3,347 \\
FM1     & 704   & 258   & 271   \\
FM10    & 704   & 258   & 270   \\
FM100   & 481   & 184   & 206   \\
FM1000  & 482   & 183   & 209   \\
\hline
\end{tabular}
\end{table}

Each transfer learning method from Table~\ref{tab:tl_list} is applied to the 100 realizations of the pretrained RNN to generate predictions of FM1, FM100, and FM1000. All accuracy metrics are reported on the test set. 

\citet{Carlson-2007-ANM} reported $R^2$ and bias for the Nelson model for each fuel class and, for the smaller fuel classes, for observations with FMC less than $30\%$. We compare our accuracy metrics to these reported values, noting that the Nelson model was evaluated over the whole study period rather than over separate training, validation, and test sets.

\section{Results}

\subsection{Zero-Shot Transfer Learning for FM10 Results}\label{sec:zeroshot}

The RNN trained on RAWS FM10 observations with HRRR inputs was applied to the Oklahoma field study, using Mesonet weather observations of the variables listed in Table~\ref{tab:all_vars} together with the geographic variables for the Slapout station~\citep{OKMeso-2026-MSI} to predict hourly FM10 from March 26, 1996, at 23:00 through December 30, 1997, at 22:00 Central Time. These predictions were evaluated against the 1,232 gravimetric FM10 observations from the Oklahoma field study for each of the 100 RNN realizations.

Table~\ref{tab:model_comparison} reports the resulting accuracy metrics for this RNN and, for comparison, the Nelson-model metrics reported by \citet[Table 4]{Carlson-2007-ANM}, for all observations and for FM10 values filtered to less than $30\%$. Both models are evaluated on the Oklahoma field study, but the RNN is trained on separate source-task data, whereas the Nelson model was calibrated and evaluated on the Oklahoma study data itself.  

\begin{table}[htbp]
\caption[Accuracy metrics for FM10 model compared to field observations in Oklahoma.]{Accuracy metrics for FM10 Zero-Shot RNN Predictions and Nelson model~\citep[Table 4, RMSE not provided]{Carlson-2007-ANM}. Comparisons of the models are presented for all FM10 values ($n=1,232$) and for FM10 $\leq 30\%$ ($n=1,134$). RNN metrics are reported as mean $\pm$ standard deviation across 100 realizations.}
\label{tab:model_comparison}
\centering
\begin{tabular}{lcc|cc}
\multicolumn{1}{c}{} 
& \multicolumn{2}{c}{All observed FM10} 
& \multicolumn{2}{c}{Observed FM10 $\leq$ 30\%} \\
\hline
Metric 
& Nelson (hourly) & RNN 
& Nelson (hourly) & RNN \\
\hline
$R^2$ & $0.79$ & $0.56\pm0.02$ & $0.61$ & $\;\;\,0.65\pm0.06$ \\
Bias (\%)  & $-0.9$ & $0.91\pm0.52$ & $-1.1$ & $-0.74\pm0.53$ \\
RMSE (\%) & - & $6.79\pm0.16$ & - & $\;\;\,3.21\pm0.25$ \\
\hline
\end{tabular}
\end{table}

The aggregate $R^2$ over all FM10 observations is substantially lower for the RNN than for the Nelson model. We believe this is almost entirely an artifact of the FM10 fuel sensors used for source-task training. Some stations in the Synoptic data network report the manufacturer of the fuel sensor, and the majority are manufactured by FTS. These sensors appear to have a maximal response to rain, where the FM10 values do not exceed about 27$\%$. 

\begin{figure}[htbp]
\centering
\begin{minipage}[c]{0.49\textwidth}
  \centering
  \includegraphics[width=\textwidth]{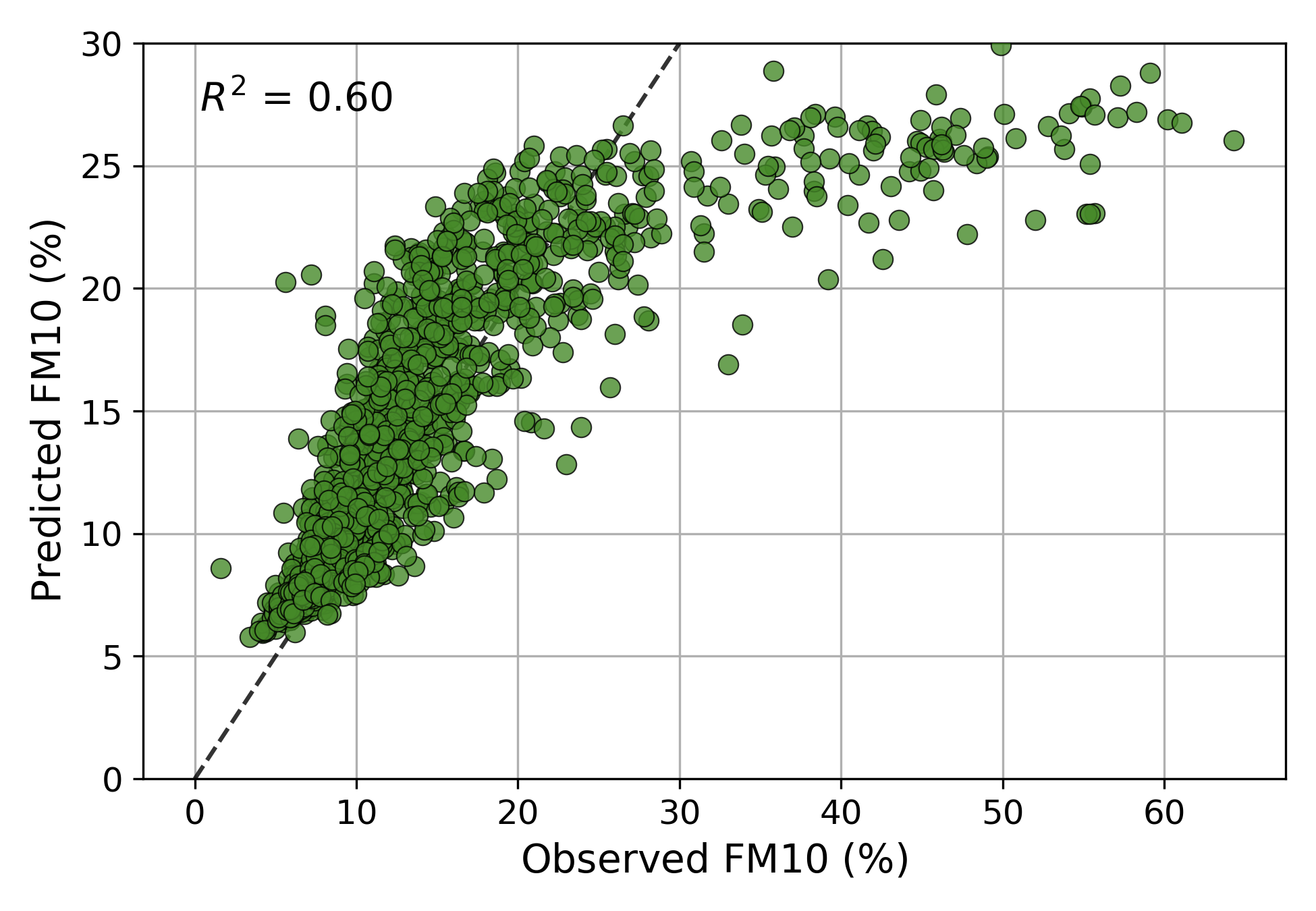}
\end{minipage}
\begin{minipage}[c]{0.49\textwidth}
  \centering
  \includegraphics[width=\textwidth]{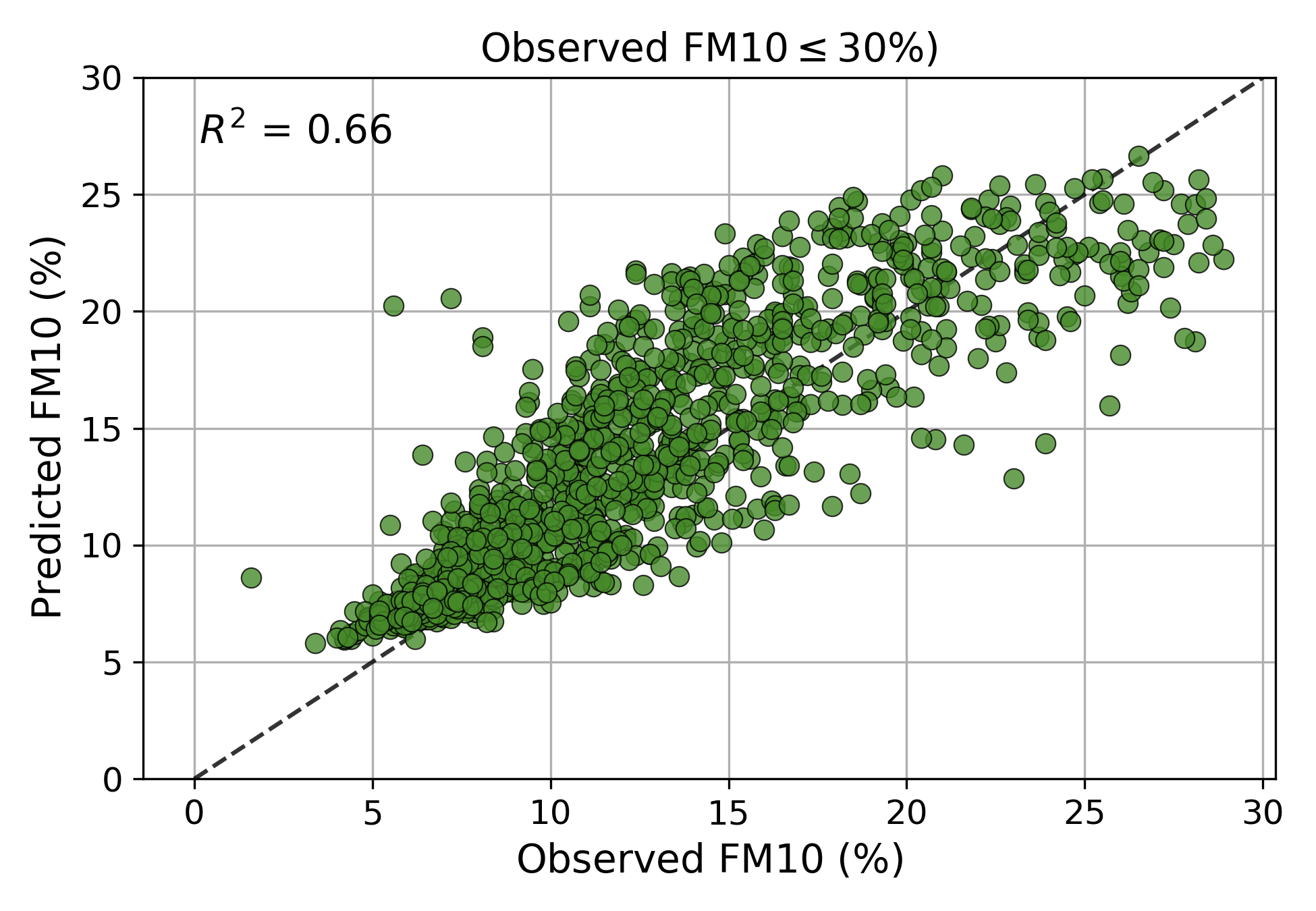}
\end{minipage}

\caption[
Observed vs Predicted FM10 for Zero-Shot Transfer RNN
]{
Observed vs Predicted FM10 for Zero-Shot Transfer RNN. The plotted RNN model corresponds to the median RMSE on the test set out of 100 realizations. Note the different x-axis ranges.
(a) Left: All FM10 observations~(n=1,232), where predictions substantially underestimate FM10 for wet fuels.
(b) Right: FM10 observations $\leq 30\%$~(n=1,134), where predictions are more accurate but still overestimate very dry fuels and underestimate wetter fuels.
}
\label{fig:fm10_scatters}
\end{figure}

Figure~\ref{fig:fm10_scatters} shows observed FM10 versus predicted FM10 from a representative zero-shot RNN.
The plotted predictions are from the model with the median RMSE in the test period over 100 realizations.
For the FM10 values less than or equal to $30\%$ used in the right panel, the median RMSE of this representative model was $3.16\%$.

\begin{figure}[htbp]
\centering
\includegraphics[width=14 cm]{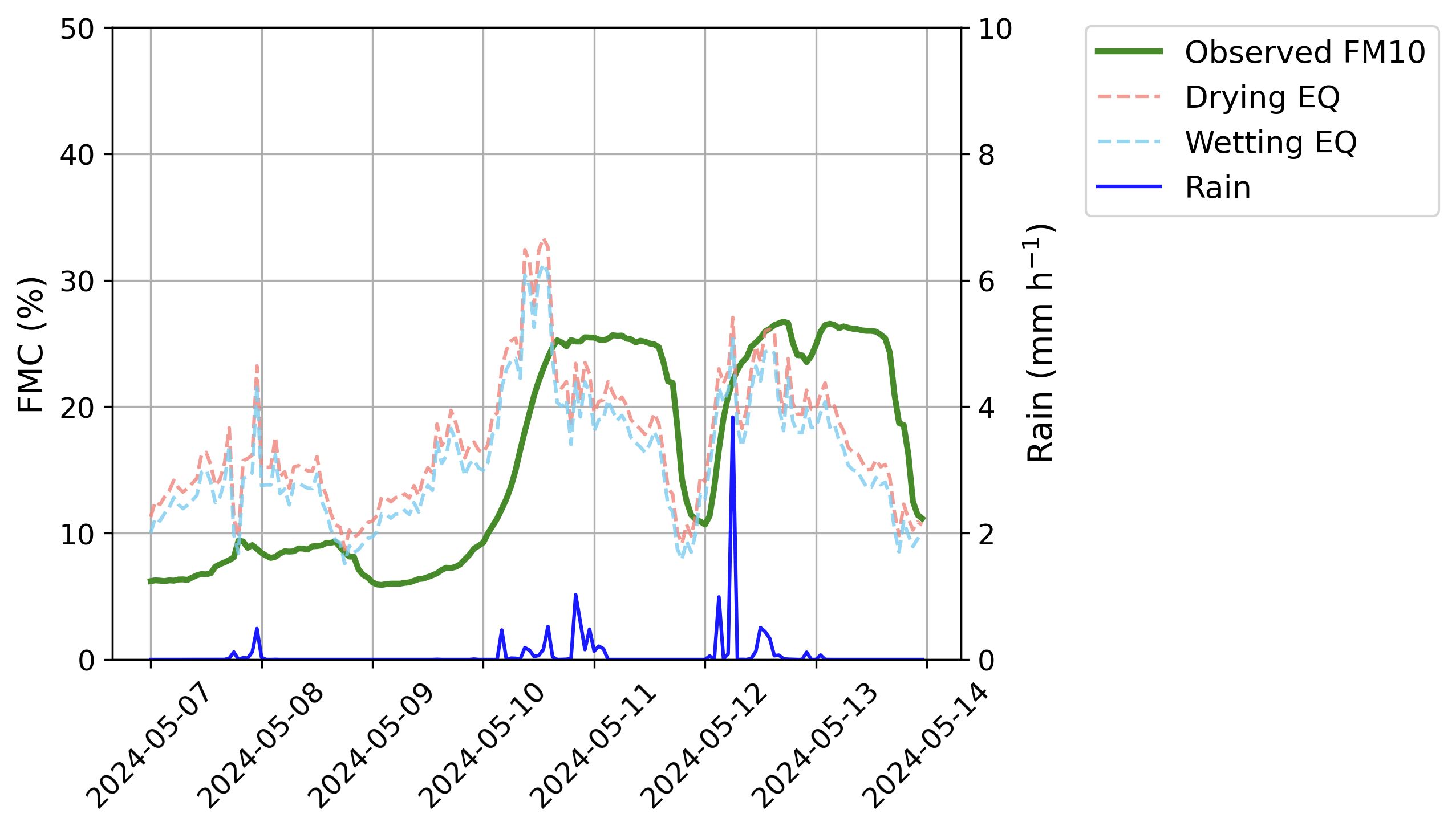}
\caption[Time series of weather and FM10 observations from a RAWS in Colorado.]{One week of FM10 sensor measurements and corresponding HRRR weather at BAWC2, southwest of Denver, Colorado. The wetting and drying equilibria are derived from the forecasted air temperature and RH. The sensor appears to have a maximal response to rain, with the FM10 reaching a maximum value of around 27$\%$.}
\label{fig:ts_BAWC2} 
\end{figure}

\begin{figure}[htbp]
\centering
\includegraphics[width=14 cm]{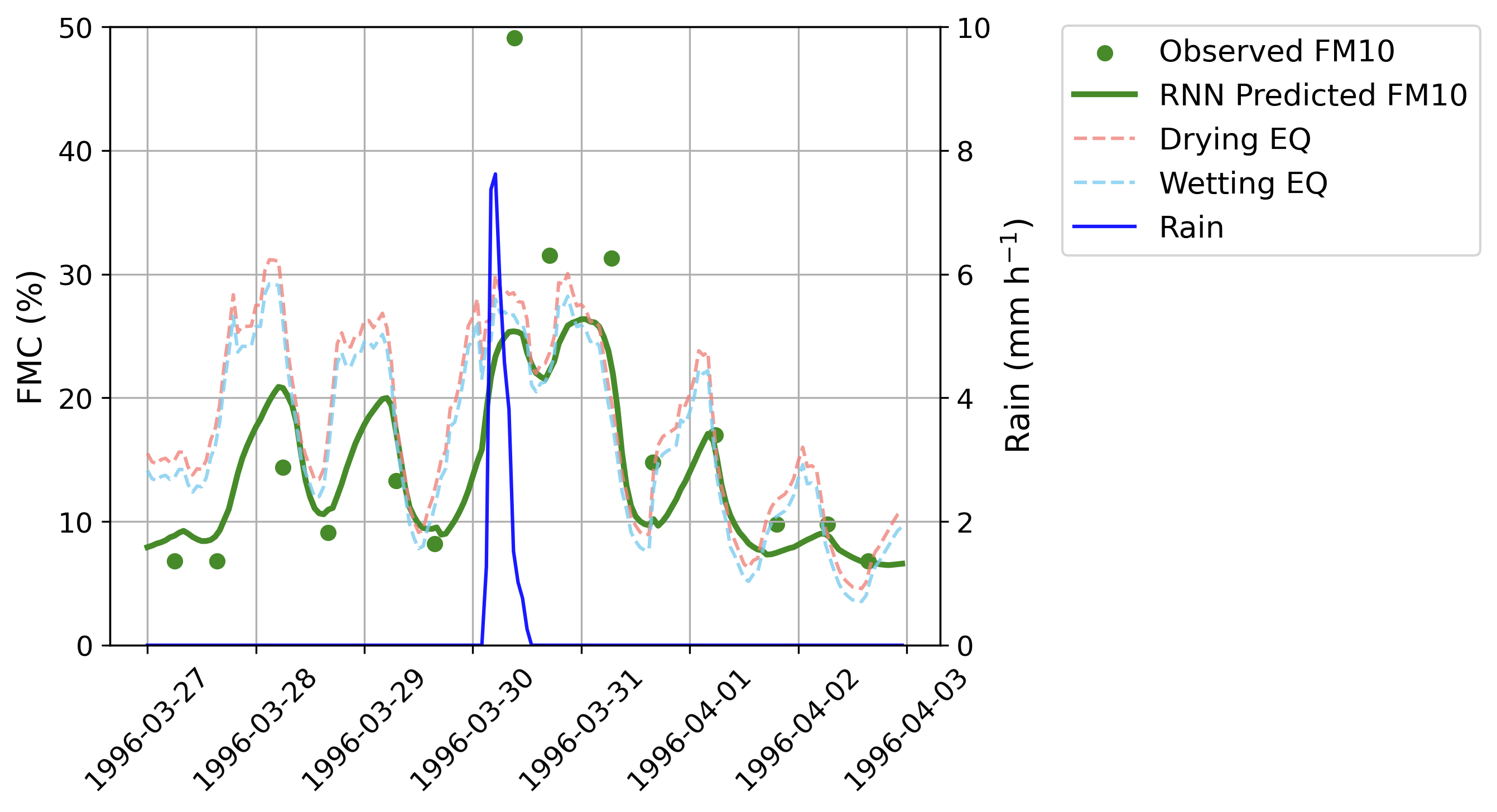}
\caption[RNN Zero-shot predictions with time series of weather and FM10 observations from Oklahoma field data.]{One week of FM10 field measurements in Oklahoma, the corresponding ground-based weather observations, and the RNN zero-shot predictions. The RNN predictions are from the model realization with the median RMSE in the test set. The RNN predictions reach a maximum of about 27$\%$, similar to the RAWS training data.}
\label{fig:ts_zeroshot} 
\end{figure}

Figure~\ref{fig:ts_BAWC2} shows one week of sensor measurements near Green Mountain, Colorado, during a period with multiple rain events forecast by HRRR, and the FM10 sensor does not exceed about 27$\%$. Figure~\ref{fig:ts_zeroshot} shows a representative Oklahoma field study example, with weather, FM10 observations, and zero-shot predictions from the realization with the median RMSE. Following a major rain event, the observed FM10 rises to nearly 50$\%$, while the RNN peaks at only about 27$\%$. These results suggest that the RNN learned the sensor saturation pattern, which produces large errors for very wet fuels. At the same time, the Oklahoma example suggests that the RNN still learned the post-rain drying behavior reasonably well, which helps explain the lower errors once FM10 returns below about $30\%$, the range that is often emphasized in FMC studies because higher values exceed the moisture of extinction for most fuels in this class~\citep[e.g.,][]{Carlson-2007-ANM,Anderson-1982-ADF}. 

For the FM10 values less than or equal to $30\%$ in Table~\ref{tab:model_comparison}, the zero-shot RNN achieved an RMSE of $3.21\pm0.25\%$ across 100 realizations.
This is consistent with the variability reported for the source RNN across realizations~\citep{Hirschi-2026-RNN}, supporting its utility in this setting.
For the same subset, the bias was negative, indicating a greater underprediction than in the fuel-sensor training data, but its magnitude remained small at $-0.74\%$ and was smaller in absolute value than for the Nelson model.

As a simple baseline, static XGBoost and linear regression models were evaluated on the same data. For all FM10 observations, they achieved slightly lower RMSE ($5.6$--$5.9\%$) than the RNN. However, for FM10 $\leq 30\%$, the RNN outperformed both models (RMSE $3.88$--$3.47\%$), indicating superior performance in the lower fuel moisture regime.

Zero-shot transfer learning for FM10 shows that RNN transfers reliably in this setting.
Although performance was poor for FM10 values above $30\%$, the zero-shot RNN in Table~\ref{tab:model_comparison} was highly accurate for FM10 $\leq 30\%$, where its metrics were comparable to those of the Nelson model.
Because the Nelson model was calibrated to the Oklahoma field study, whereas the RNN was applied there without fine-tuning after pretraining on different data sources, this result is encouraging and suggests that the RNN may provide more accurate FMC forecasts across locations and times beyond this study.

\subsection{FM1 Transfer Learning Results}

\begin{figure}[htbp]
\centering
\includegraphics[width=14 cm]{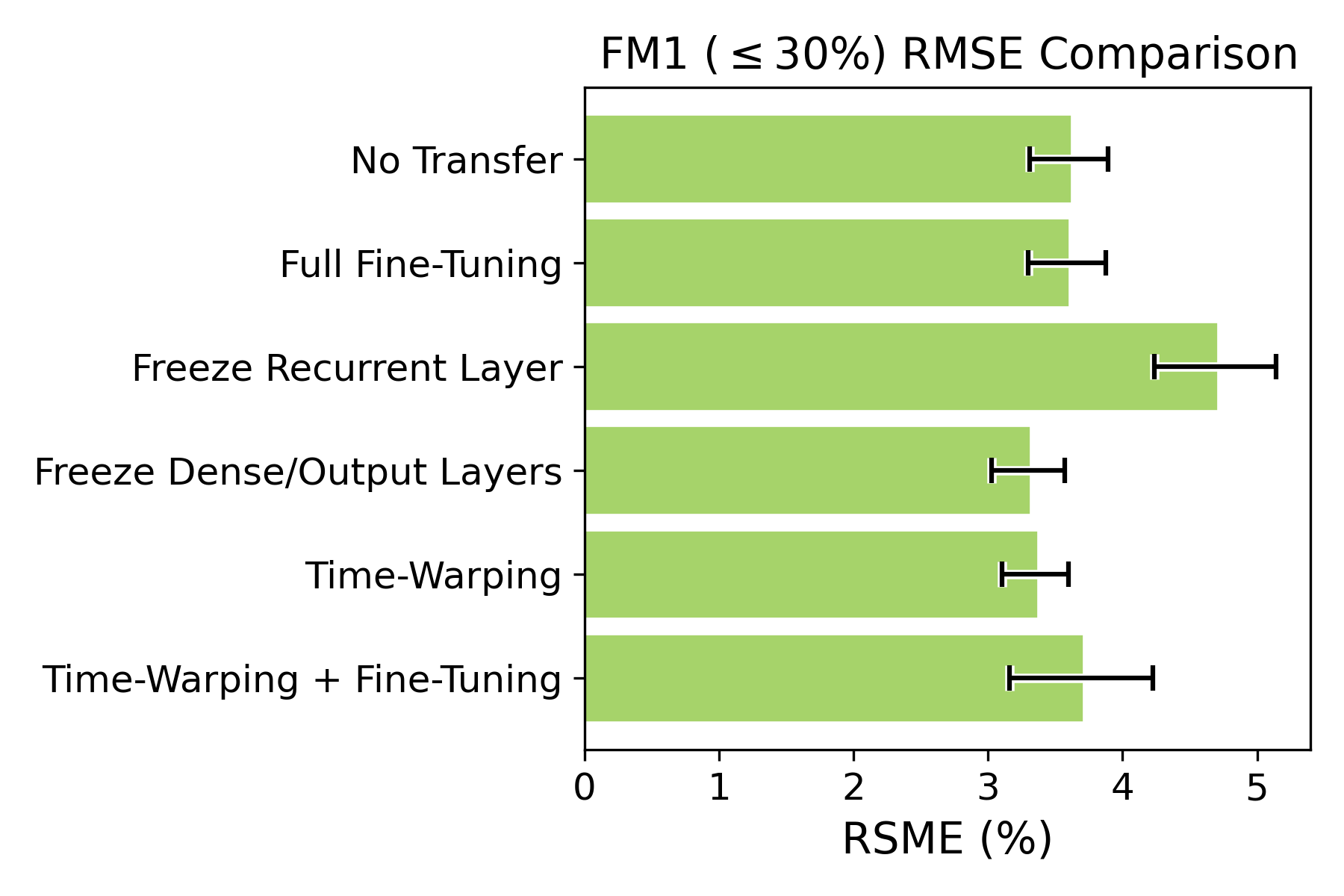}
\caption[FM1 ($\leq 30\%$) RMSE by Transfer Learning Method.
]{FM1 RMSE comparison for different transfer learning methods with FM1 filtered to ($\leq 30\%$). Bars represent the mean RMSE when predicting FM1 on the test set over 100 replications, and brackets show $\pm$ 1 standard deviation of the RMSE. }
\label{fig:fm1_30_rmse} 
\end{figure}

For FM1, 271 observations were collected in the Oklahoma test set, of which 247 had values less than or equal to $30\%$.
Table~\ref{tab:fm1_all} shows the accuracy metrics for the different transfer learning approaches in the entire FM1 test set, and Table~\ref{tab:fm1_le30} shows the same comparison for FM1 values less than or equal to $30\%$. Figure~\ref{fig:fm1_30_rmse} shows a visual comparison of the RMSE statistics for the transfer learning methods with FM1 filtered to values less than or equal to $30\%$.

\begin{table}[htbp]
\centering
\caption[FM1 Accuracy Metrics - Transfer Learning Comparison.]{Accuracy metrics for FM1 RNN predictions compared to field observations in Oklahoma for different transfer learning methods. Metrics are reported as mean $\pm$ standard deviation across 100 realizations~(n=271), except for the Nelson model baseline.}
\label{tab:fm1_all}
\begin{tabular}{lccc}
\hline
Method & $R^2$ & Bias (\%) & RMSE (\%) \\
\hline
No Transfer & $0.78 \pm 0.03$ & $0.84 \pm 0.71$ & $6.31 \pm 0.40$ \\
Full Fine-Tuning & $0.84 \pm 0.02$ & $2.24 \pm 0.53$ & $5.31 \pm 0.38$ \\
Freeze Recurrent Layer & $0.79 \pm 0.02$ & $1.23 \pm 0.76$ & $6.06 \pm 0.36$ \\
Freeze Dense/Output Layers & $0.84 \pm 0.03$ & $2.19 \pm 0.46$ & $5.39 \pm 0.43$ \\
Time-Warping & $0.41 \pm 0.03$ & $2.46 \pm 0.42$ & $10.29 \pm 0.23$ \\
Time-Warping + Fine-Tuning & $0.84 \pm 0.05$ & $1.11 \pm 1.04$ & $5.26 \pm 0.71$ \\
\hline\hline
Nelson Model* & $0.64$ & $-2.7$ & - \\
\hline
\end{tabular}

\vspace{0.5em}
\footnotesize{
* Nelson model metrics were reported over the full study period rather than the test subset used for the transfer learning methods.
}
\end{table}

Across all 271 FM1 observations in Table~\ref{tab:fm1_all}, the RNN-based methods have RMSE values above $5\%$, largely due to the difficulty of modeling very wet fuels.
The Time-Warping method without fine-tuning performs worst in Table~\ref{tab:fm1_all}, with RMSE above $10\%$ and a positive bias of $2.46\%$, because it starts from an FM10 RNN trained on sensors with a limited rain response and adjusts only the time-warping parameters.
Fine-tuning reduces this error substantially, with Time-Warping plus Fine-Tuning achieving the lowest mean RMSE ($5.26\%$) and tying for the best mean $R^2$ ($0.84$), with variability across realizations similar to that of Full Fine-Tuning and Freeze Dense/Output Layers.

\begin{table}[htbp]
\centering
\caption[FM1 Accuracy Metrics ($\leq 30\%$) - Transfer Learning Comparison.]{Accuracy metrics for FM1 RNN predictions compared to field observations in Oklahoma for different transfer learning methods. Metrics are reported as mean $\pm$ standard deviation across 100 realizations. FM1 values are filtered $\leq 30\%$~(n=247).}
\label{tab:fm1_le30}
\begin{tabular}{lccc}
\hline
Method & $R^2$ & Bias (\%) & RMSE (\%) \\
\hline
No Transfer & $0.66 \pm 0.06$ & $-0.30 \pm 0.62$ & $3.60 \pm 0.29$ \\
Full Fine-Tuning & $0.67 \pm 0.05$ & $1.43 \pm 0.49$ & $3.59 \pm 0.29$ \\
Freeze Recurrent Layer & $0.43 \pm 0.11$ & $0.30 \pm 0.74$ & $4.69 \pm 0.45$ \\
Freeze Dense/Output Layers & $0.72 \pm 0.05$ & $1.40 \pm 0.40$ & $3.30 \pm 0.27$ \\
Time-Warping & $0.71 \pm 0.04$ & $-0.28 \pm 0.43$ & $3.35 \pm 0.25$ \\
Time-Warping + Fine-Tuning & $0.64 \pm 0.11$ & $0.62 \pm 0.92$ & $3.69 \pm 0.53$ \\
\hline\hline
Nelson Model* & $0.46$ & $-3.6$ & - \\
\hline
\end{tabular}

\vspace{0.5em}
\footnotesize{
* Nelson model metrics were reported over the full study period rather than the test subset used for the transfer learning methods.
}
\end{table}

For the FM1 values less than or equal to $30\%$ in Table~\ref{tab:fm1_le30}, RMSE decreases substantially for all RNN-based methods, while most $R^2$ values decrease because removing very wet observations reduces the variance of the target.
For the same filtered subset, the Nelson model had the second worst $R^2$ of $0.43$ and the largest bias at $-3.6\%$.

When FM1 values above $30\%$ were excluded, the Time-Warping method in Table~\ref{tab:fm1_le30} performed much better than in Table~\ref{tab:fm1_all}.
Its $R^2$ was among the best, its bias was consistent with zero, and its RMSE was the second best and within the variability across realizations of the most accurate method.
It also outperformed the Nelson model in $R^2$, with the caveat that the comparison is not exact.

Interestingly, for the FM1 values less than or equal to $30\%$ in Table~\ref{tab:fm1_le30}, fine-tuning did not improve the Time-Warping method.
One possible explanation is that fine-tuning reduced errors for very wet fuels during training at the expense of accuracy for drier fuels. 

The other transfer learning methods in Tables~\ref{tab:fm1_all} and~\ref{tab:fm1_le30} also help to clarify which aspects of the problem are generalized from FM10.
Full Fine-Tuning did not substantially improve on the No Transfer baseline.
One possible explanation is that full fine-tuning prioritizes reducing large errors during rain events, which may come at the expense of accuracy in the drier regime.
Freezing the recurrent layer performed worst, supporting the need to adapt the temporal dynamics when transferring from the FM10 source task.
Freezing dense and output layers was the most accurate method, but the Time-Warping method remained competitive in both $R^2$ and RMSE while achieving better bias.

The test set for FM1 less than or equal to $30\%$ in Table~\ref{tab:fm1_le30} contained only 247 observations from a single location over four months, so it is difficult to assess how well the more fine-tuned methods would generalize.
By contrast, the Time-Warping method fit only two parameters, modifying 128 of more than 21,000 trainable parameters, and its FM1 accuracy was consistent with that of the pretrained FM10 RNN evaluated on a much larger multi-site dataset.

\begin{figure}[htbp]
\centering
\begin{minipage}[c]{0.49\textwidth}
  \centering
  \includegraphics[width=\textwidth]{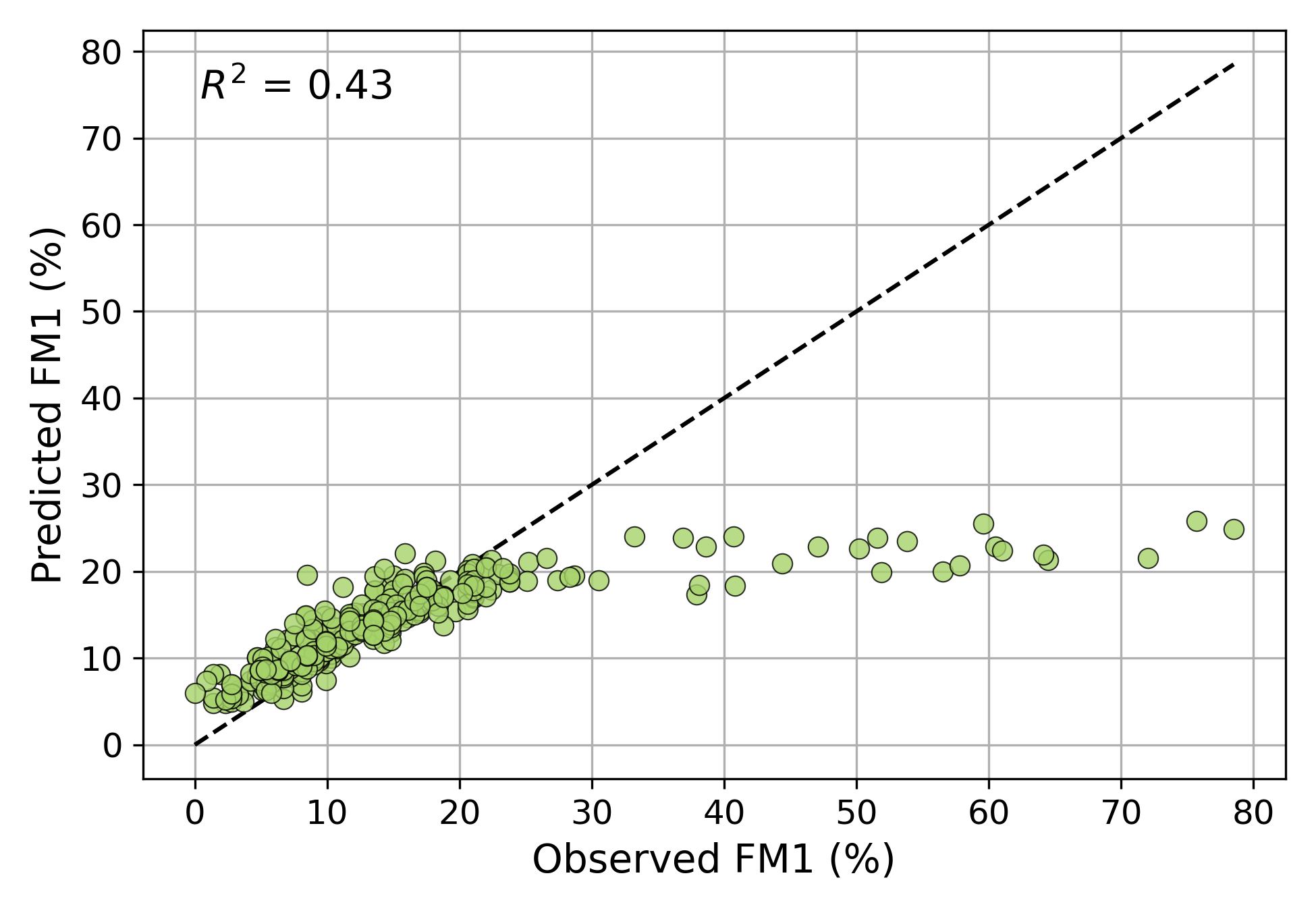}
\end{minipage}
\begin{minipage}[c]{0.49\textwidth}
  \centering
  \includegraphics[width=\textwidth]{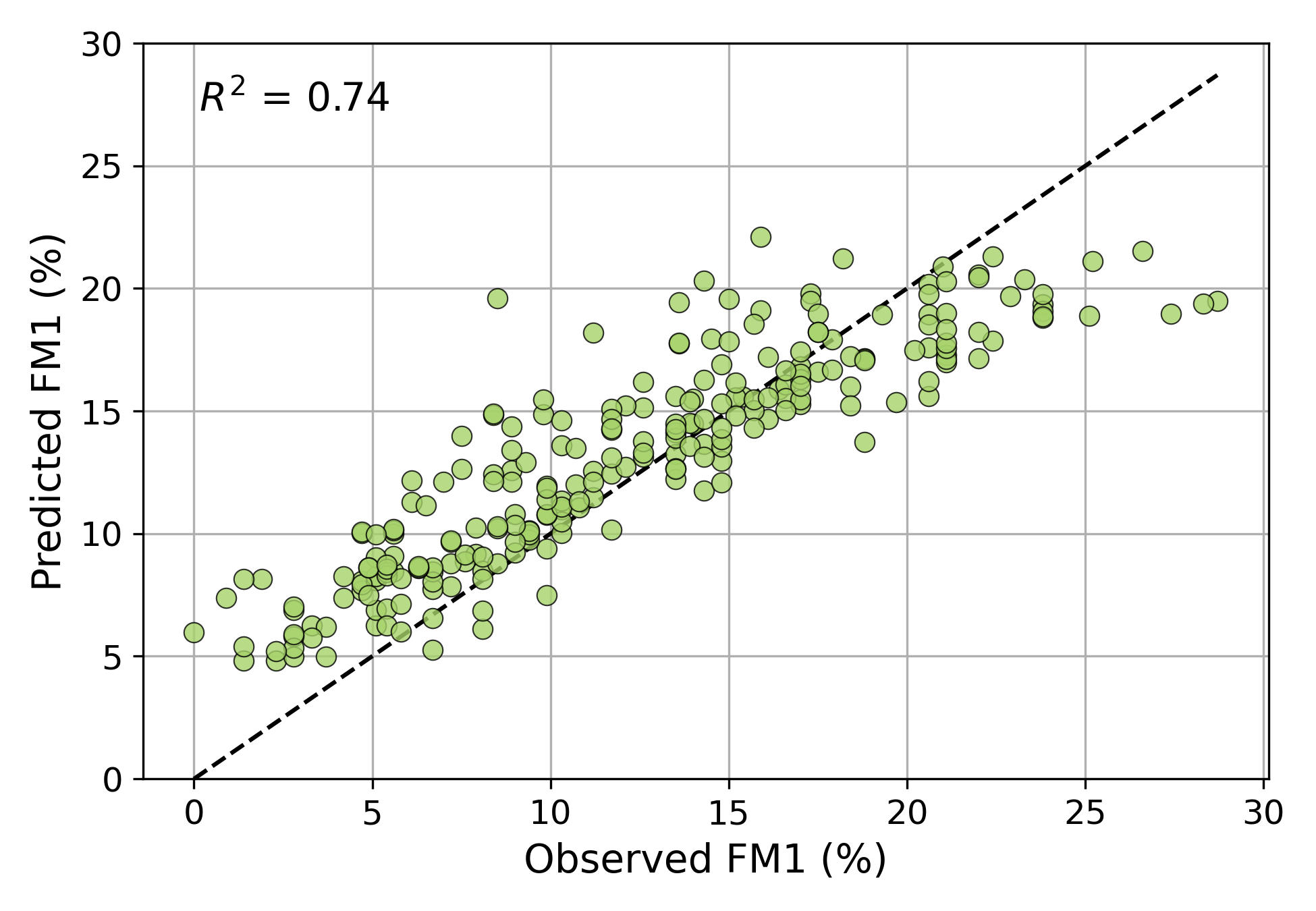}
\end{minipage}

\caption[
Observed vs Predicted FM1 for Time-Warped RNN
]{
Observed vs Predicted FM1 for Time-Warped RNN. The plotted RNN model corresponds to the median RMSE on the test set out of 100 realizations.  Note the different x-axis ranges.
(a) Left: All FM1 observations~(n=271), where predictions substantially underestimate FM1 for wet fuels.
(b) Right: FM1 observations $\leq 30\%$~(n=247), where predictions are more accurate but still overestimate very dry fuels and underestimate wetter fuels.
}
\label{fig:fm1_scatters}
\end{figure}

Figure~\ref{fig:fm1_scatters} shows observed versus predicted FM1 for a representative Time-Warping RNN whose RMSE was closest to the median in the 100 realizations.
For FM1$\leq30\%$, shown in the right panel, the RNN tends to overestimate drier fuels and underestimate wetter fuels, but overall agreement with observations remains strong.

\subsection{FM100 Transfer Learning Results}

\begin{figure}[htbp]
\centering
\includegraphics[width=14 cm]{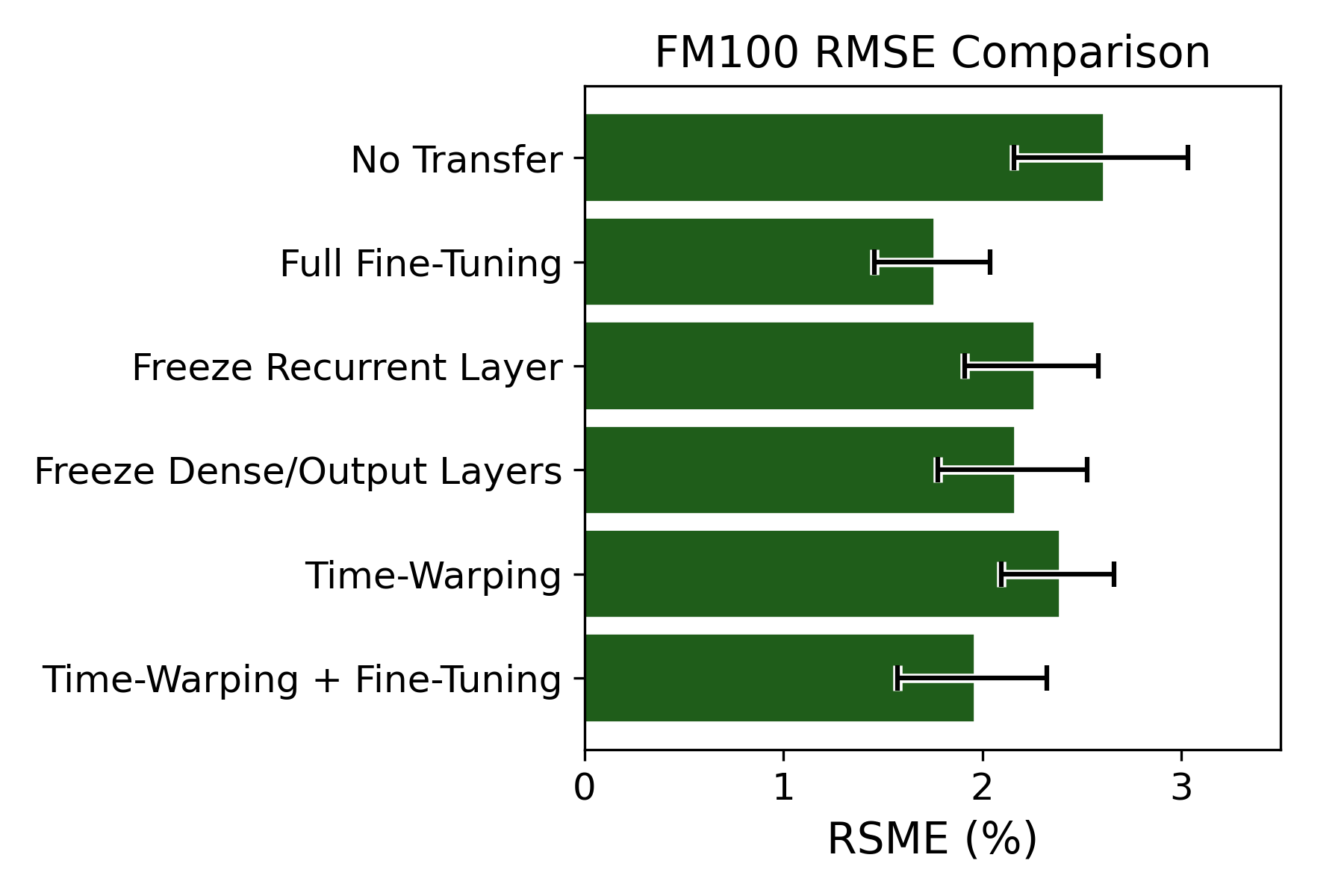}
\caption[FM100 RMSE by Transfer Learning Method.
]{FM100 RMSE comparison for different transfer learning methods. Bars represent the mean RMSE when predicting FM100 on the test set over 100 replications, and brackets show $\pm$ 1 standard deviation of the RMSE. }
\label{fig:fm100_rmse} 
\end{figure}

For FM100, 206 observations were collected in the Oklahoma test set, and no filtering was applied for wet fuels.
Table~\ref{tab:fm100_all} shows the accuracy metrics for the different transfer learning approaches transferred from the pretrained FM10 model, together with the corresponding metrics for the Nelson model. Figure~\ref{fig:fm100_rmse} shows a visual comparison of the RMSE statistics for the transfer learning methods.

\begin{table}[htbp]
\centering
\caption[FM100 Accuracy Metrics - Transfer Learning Comparison.]{Accuracy metrics for FM100 RNN predictions compared to field observations in Oklahoma for different transfer learning methods~(n=206). Metrics are reported as mean $\pm$ standard deviation across 100 realizations.}
\label{tab:fm100_all}
\begin{tabular}{lccc}
\hline
Method & $R^2$ & Bias (\%) & RMSE (\%) \\
\hline
No Transfer & $0.64 \pm 0.13$ & $-0.19 \pm 0.33$ & $2.60 \pm 0.44$ \\
Full Fine-Tuning & $0.84 \pm 0.06$ & $-0.41 \pm 0.41$ & $1.75 \pm 0.29$ \\
Freeze Recurrent Layer & $0.73 \pm 0.08$ & $-0.34 \pm 0.33$ & $2.25\pm 0.34$ \\
Freeze Dense/Output Layers & $0.75 \pm 0.09$ & $0.12 \pm 0.44$ & $2.15 \pm 0.38$ \\
Time-Warping & $0.70 \pm 0.08$ & $-0.33 \pm 0.72$ & $2.38 \pm 0.28$ \\
Time-Warping + Fine-Tuning & $0.78 \pm 0.10$ & $-0.64 \pm 0.56$ & $2.00 \pm 0.40$ \\
\hline\hline
Nelson Model* & $0.75$ & $-2.1$ & - \\
\hline
\end{tabular}

\vspace{0.5em}
\footnotesize{
* Nelson model metrics were reported over the full study period rather than the test subset used for the transfer learning methods.
}
\end{table}

All methods in Table~\ref{tab:fm100_all} were more accurate for FM100 than for FM1, likely because FM100 has lower variability and slower dynamics.
The RNN-based methods in Table~\ref{tab:fm100_all} also appear to be systematically unbiased.
Although Full Fine-Tuning and Time-Warping + Fine-Tuning achieved the lowest RMSE values, the Time-Warping method remained competitive and improved modestly over the No Transfer baseline.
As in the FM1 case, it achieved comparable accuracy to the standard transfer learning baselines while modifying only a small fraction of the network parameters.

\begin{figure}[htbp]
\centering
\includegraphics[width=14 cm]{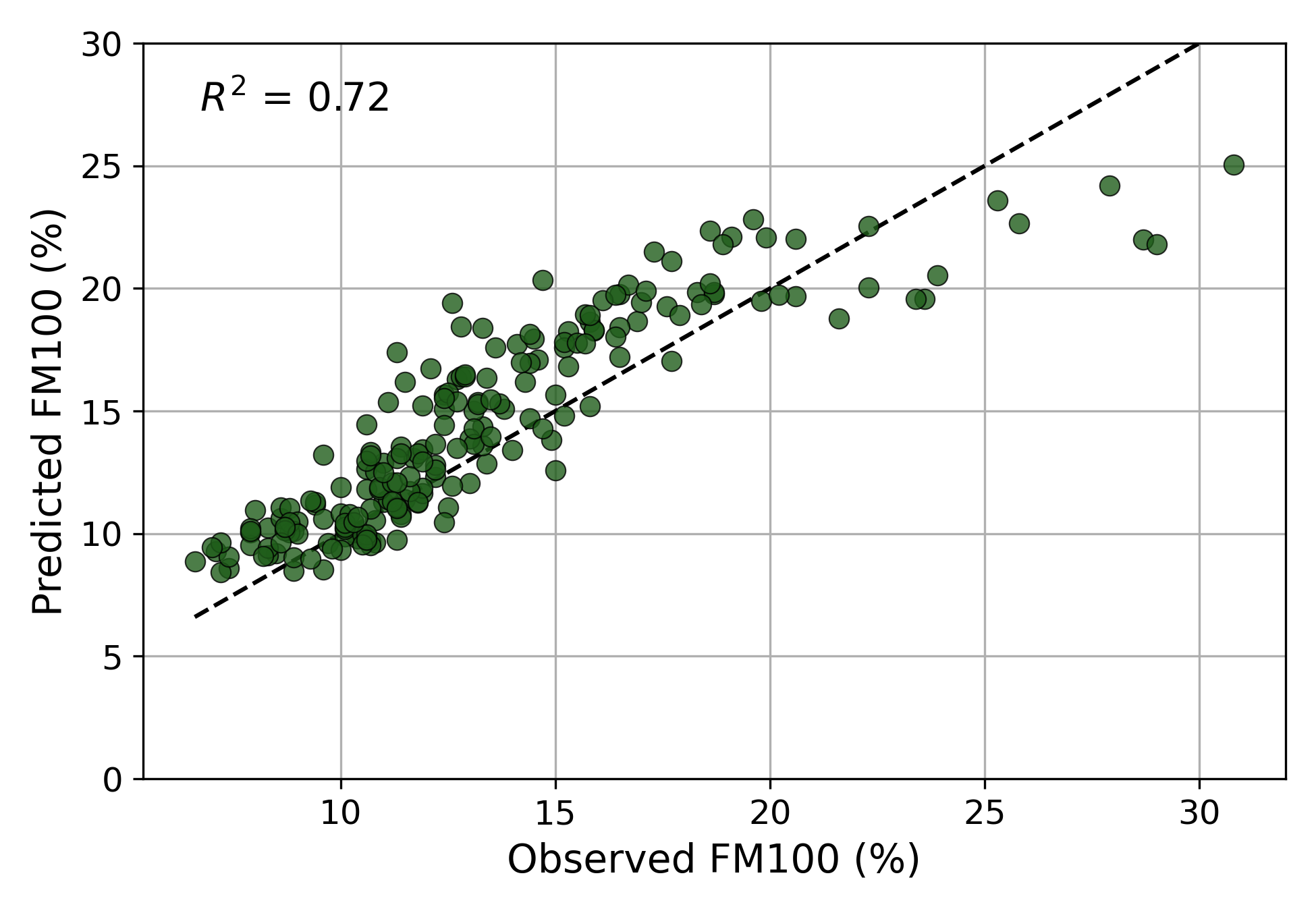}
\caption[Observed vs Predicted FM100 for Time-Warped RNN
]{Observed vs Predicted FM100 for Time-Warped RNN. The plot includes all FM100 observations in the test set~(n=206). The plotted RNN model is the set of weights that corresponds to the median RMSE on the test set out of 100 realizations. }
\label{fig:fm100_scatter} 
\end{figure}

Figure~\ref{fig:fm100_scatter} shows observed versus predicted FM100 for a representative Time-Warping RNN whose RMSE was closest to the median across the 100 realizations.
As in the FM1 case, the RNN tends to overestimate drier fuels and underestimate wetter fuels, but overall agreement with the observations remains strong.

\subsection{FM1000 Transfer Learning Results}

\begin{figure}[htbp]
\centering
\includegraphics[width=14 cm]{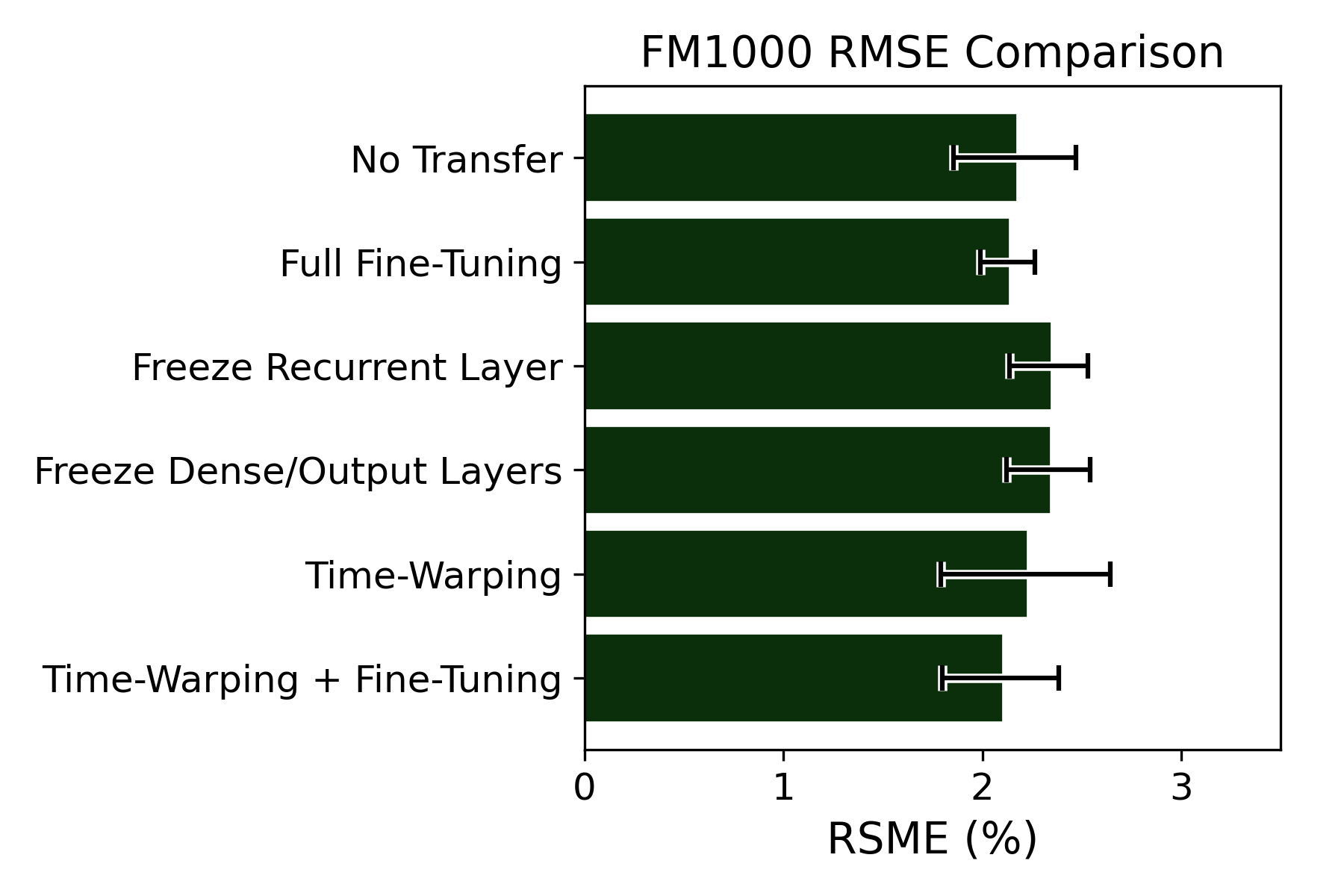}
\caption[FM1000 RMSE by Transfer Learning Method.
]{FM1000 RMSE comparison for different transfer learning methods. Bars represent the mean RMSE when predicting FM1000 on the test set over 100 replications, and brackets show $\pm$ 1 standard deviation of the RMSE. }
\label{fig:fm1000_rmse} 
\end{figure}

For FM1000, 209 observations were collected in the Oklahoma test set, and no filtering was applied for wet fuels.
Table~\ref{tab:fm1000_all} shows the accuracy metrics for the different transfer learning approaches transferred from the pretrained FM10 model, along with the corresponding metrics for the Nelson model. Figure~\ref{fig:fm1000_rmse} shows a visual comparison of the RMSE statistics for the transfer learning methods.

\begin{table}[htbp]
\centering
\caption[FM1000 Accuracy Metrics - Transfer Learning Comparison.]{Accuracy metrics for FM1000 RNN predictions compared to field observations in Oklahoma for different transfer learning methods~(n=209). Metrics are reported as mean $\pm$ standard deviation across 100 realizations.}
\label{tab:fm1000_all}
\begin{tabular}{lccc}
\hline
Method & $R^2$ & Bias (\%) & RMSE (\%) \\
\hline
No Transfer & $0.57 \pm 0.12$ & $-0.50 \pm 0.29$ & $2.16 \pm 0.31$ \\
Full Fine-Tuning & $0.59 \pm 0.05$ & $-0.92 \pm 0.21$ & $2.13 \pm 0.14$ \\
Freeze Recurrent Layer & $0.50 \pm 0.09$ & $-0.42 \pm 0.55$ & $2.33\pm 0.20$ \\
Freeze Dense/Output Layers & $0.50 \pm 0.09$ & $0.65 \pm 0.70$ & $2.33 \pm 0.21$ \\
Time-Warping & $0.54 \pm 0.18$ & $-0.81 \pm 0.71$ & $2.22 \pm 0.43$ \\
Time-Warping + Fine-Tuning & $0.65 \pm 0.10$ & $-0.75 \pm 0.36$ & $1.94 \pm 0.27$ \\
\hline\hline
Nelson Model* & $0.56$ & $-1.0$ & - \\
\hline
\end{tabular}

\vspace{0.5em}
\footnotesize{
* Nelson model metrics were reported over the full study period rather than the test subset used for the transfer learning methods.
}
\end{table}

The prediction accuracy for FM1000 in Table~\ref{tab:fm1000_all} was similar to that of FM100 and better than that of FM1.
Although the $R^2$ values were lower than for FM100, the RMSE values were similar.
In Table~\ref{tab:fm1000_all}, the Time-Warping method remained competitive with the most accurate transfer learning method, but its bias suggests a tendency to overpredict FM1000.
Its $R^2$ was also consistent with that of the Nelson model, with the usual caveat that the comparison is not exact.

\begin{figure}[htbp]
\centering
\includegraphics[width=14 cm]{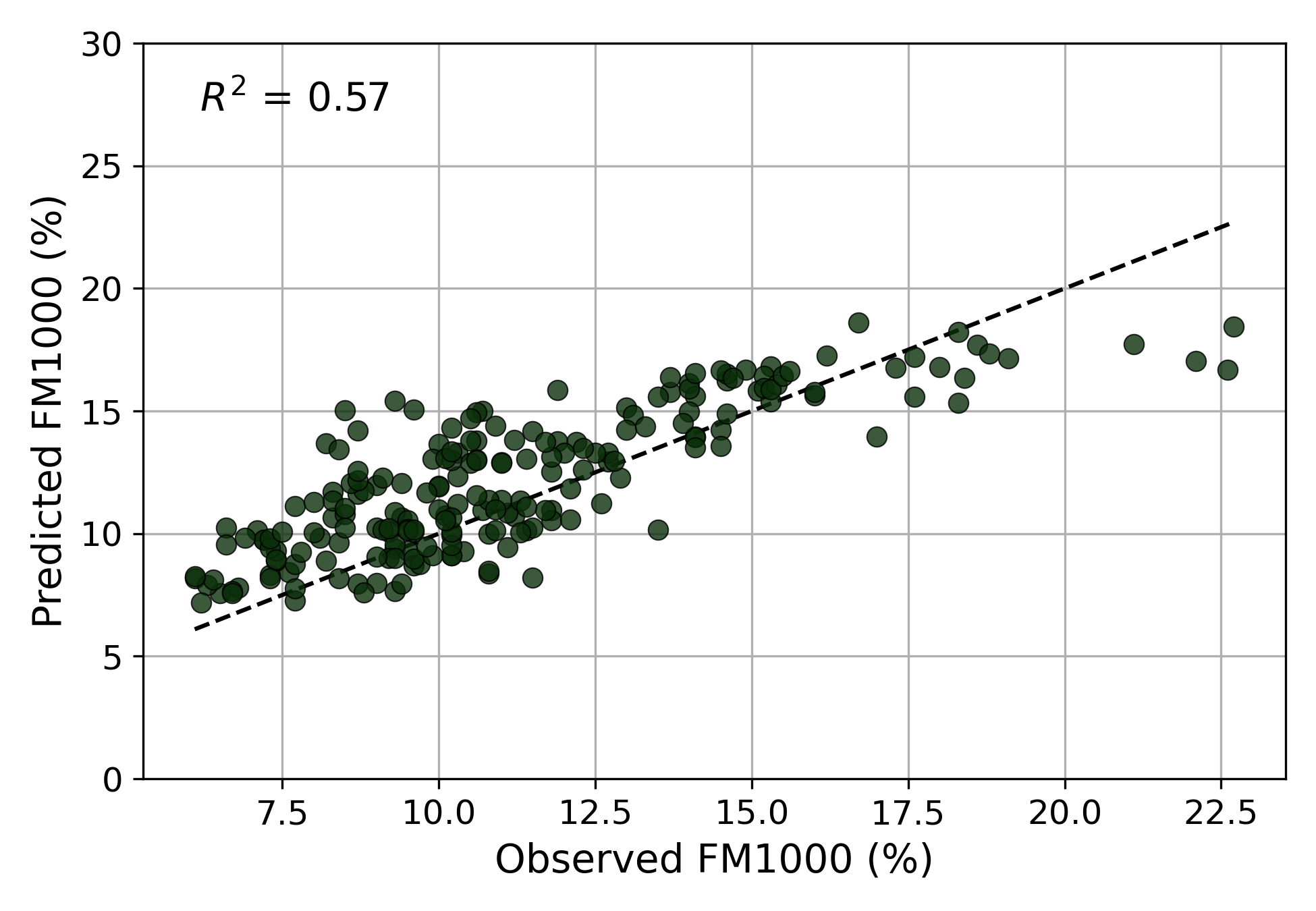}
\caption[Observed vs Predicted FM1000 for Time-Warped RNN
]{Observed vs Predicted FM1000 for Time-Warped RNN. The plot includes all FM1000 observations in the test set~(n=209). The plotted RNN model is the set of weights that corresponds to the median RMSE on the test set out of 100 realizations. }
\label{fig:fm1000_scatter} 
\end{figure}

Figure~\ref{fig:fm1000_scatter} shows observed versus predicted FM1000 for a representative Time-Warping RNN whose RMSE was closest to the median in the 100 realizations.
As in the FM1 case, the RNN tends to overestimate drier fuels and underestimate wetter fuels, although the observations above $15\%$ are sparse.
Overall, agreement with the observations is moderately strong.

\clearpage

\subsection{Time-Warping Effect on Learned Dynamics Results}

This section analyzes the learned dynamics of the RNN after applying the Time-Warping method without fine-tuning.
We use 100 realizations in which only the time-warping parameters are fit while all other model parameters are held fixed.
To assess whether the dynamics were changed, we examined the time series structure of the FM1, FM100, and FM1000 predictions during the Oklahoma test period.
The FM1 predictions change more rapidly over time than the FM100 and FM1000 predictions, while the two larger fuel classes show similar behavior.

\begin{figure}[htbp]
\centering
\includegraphics[width=14 cm]{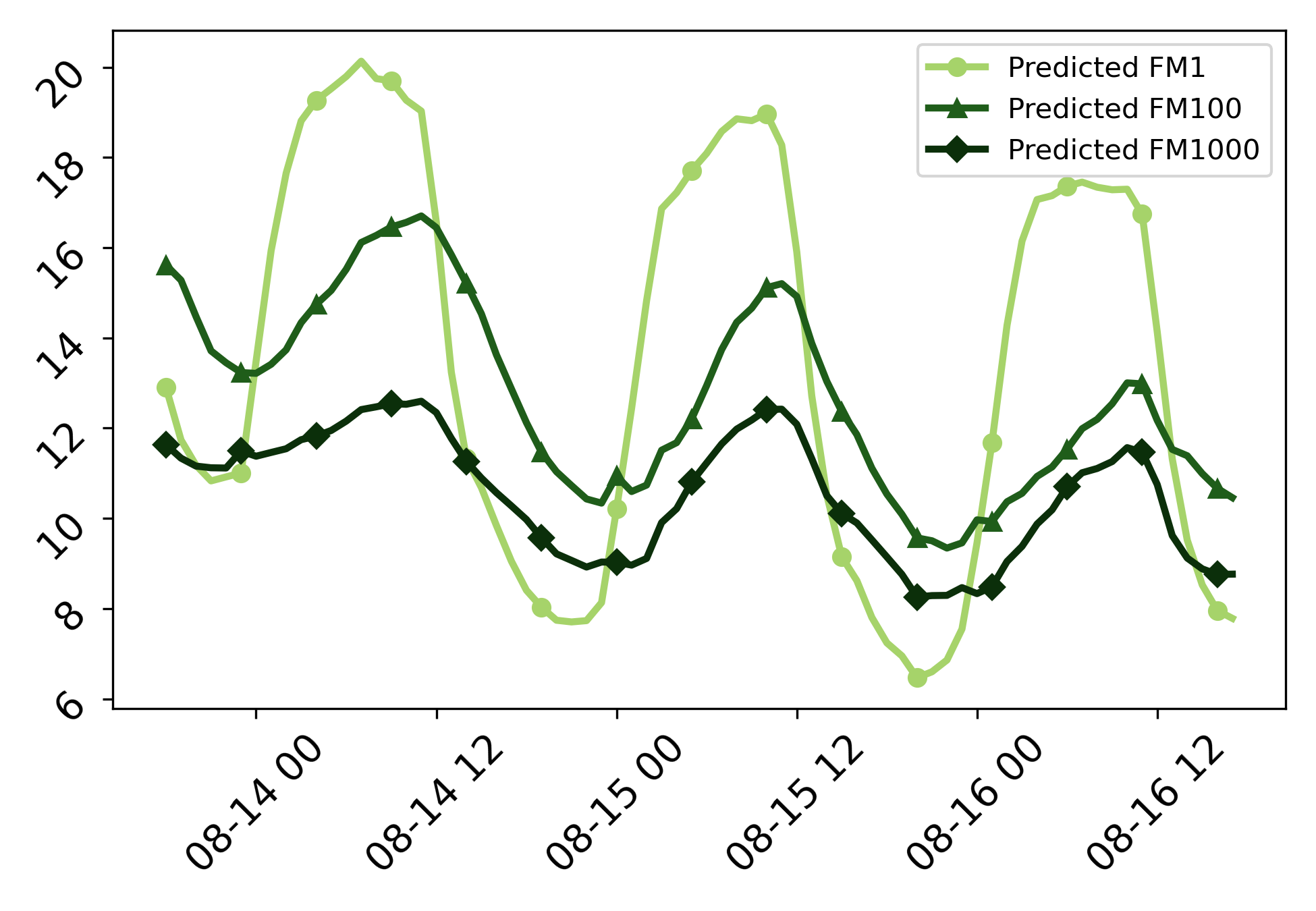}
\caption[Example Time Series of Predictions for FM1, FM100, and FM1000 from the Time-Warped RNN
]{Example Time Series of Predictions for FM1, FM100, and FM1000 from the Time-Warped RNN, with no fine-tuning. The time period plotting is 72 hours during the first week of the test period in August 1997. The predictions are from the models that had the median RMSE across the 100 realizations. }
\label{fig:tscales} 
\end{figure}

Figure~\ref{fig:tscales} shows 72 hours of Time-Warping predictions for FM1, FM100, and FM1000 from representative models with median RMSE in the 100 realizations.
The FM1 predictions vary much more throughout the diurnal cycle, consistent with a faster response to environmental changes.
The FM100 and FM1000 predictions vary more slowly and show similar dynamics, although FM100 remains wetter on average over this period.

\begin{figure}[htbp]
\centering
\begin{minipage}{0.48\textwidth}
  \includegraphics[width=\linewidth,keepaspectratio]{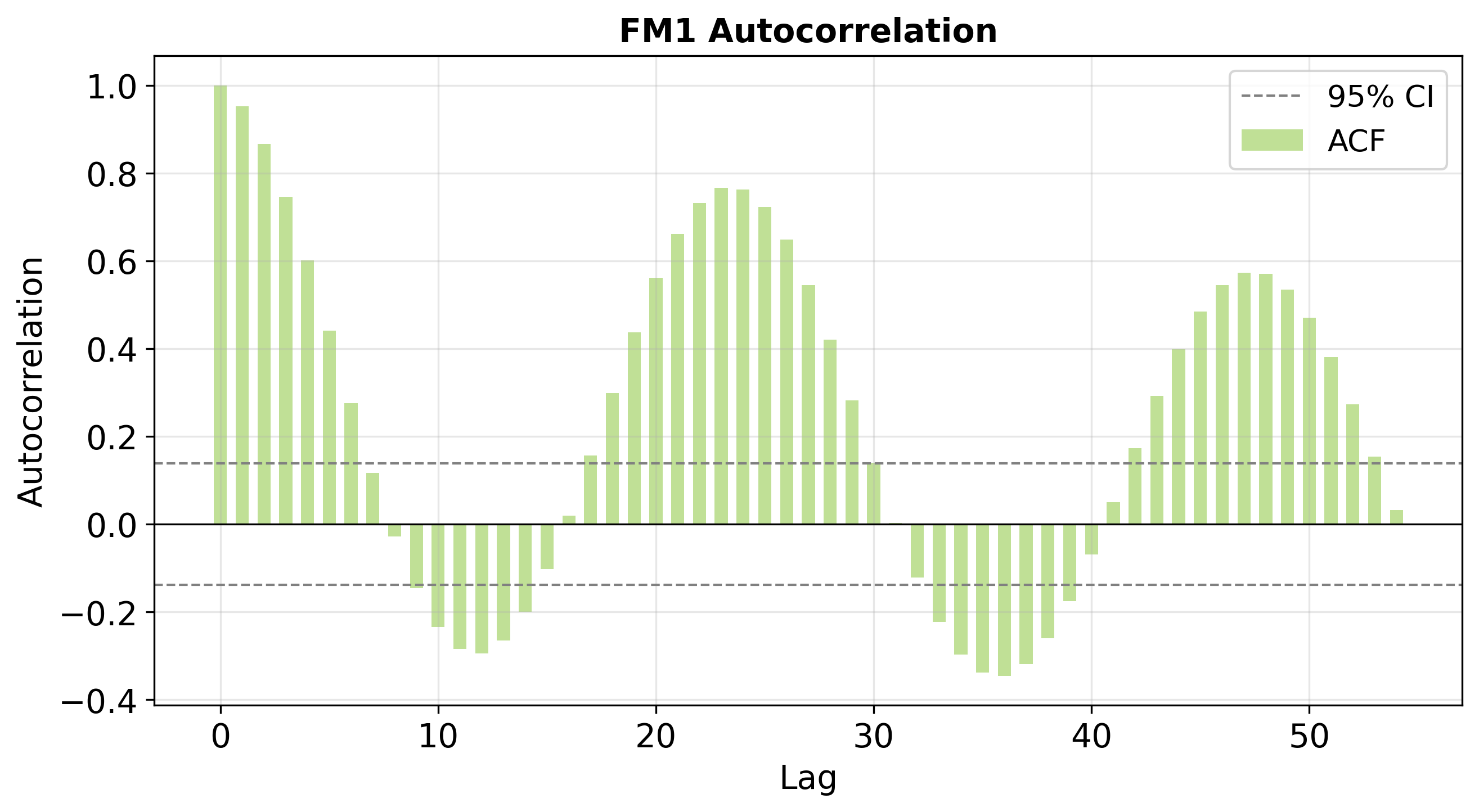}
\end{minipage}
\begin{minipage}{0.48\textwidth}
  \includegraphics[width=\linewidth,keepaspectratio]{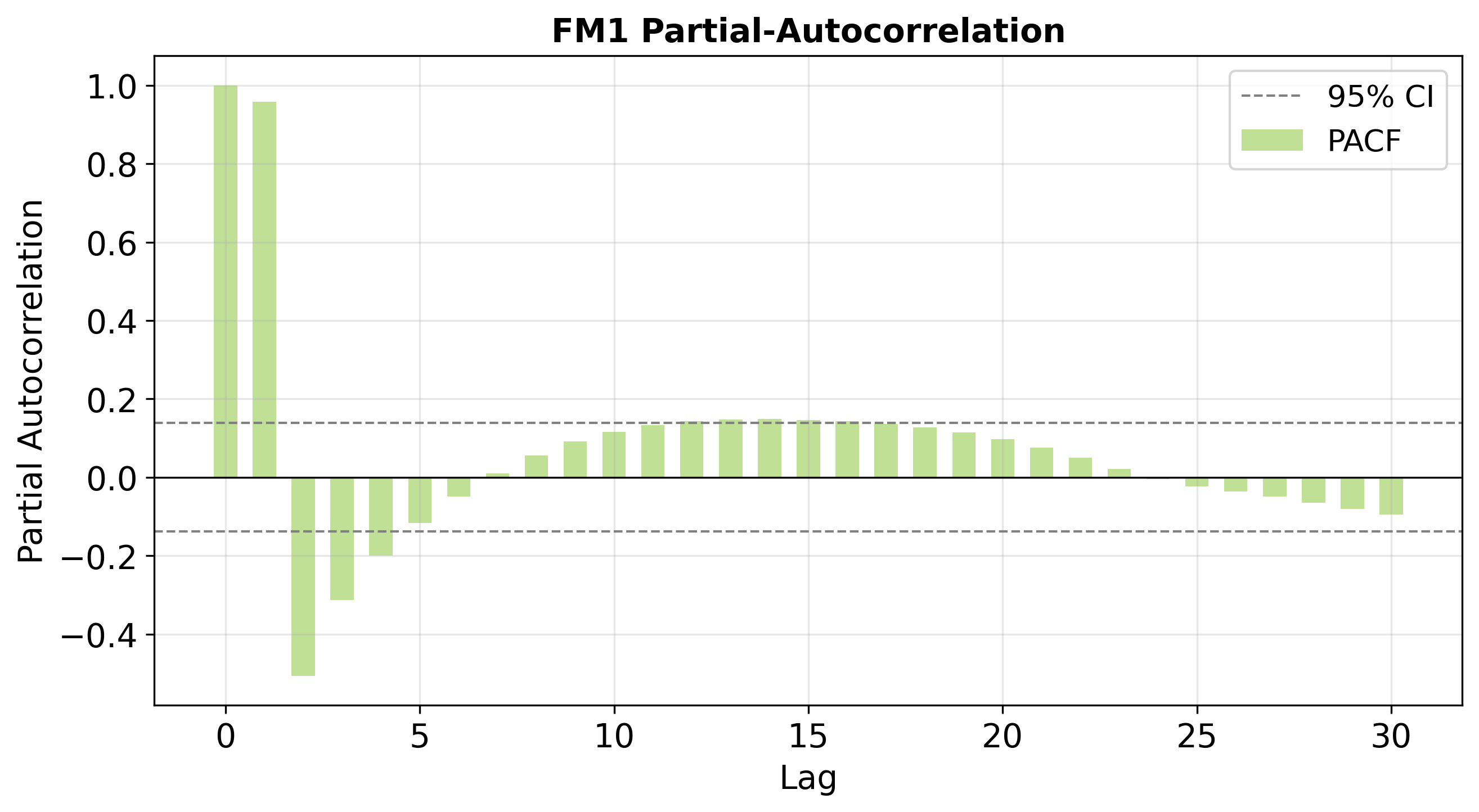}
\end{minipage}

\begin{minipage}{0.48\textwidth}
  \includegraphics[width=\linewidth,keepaspectratio]{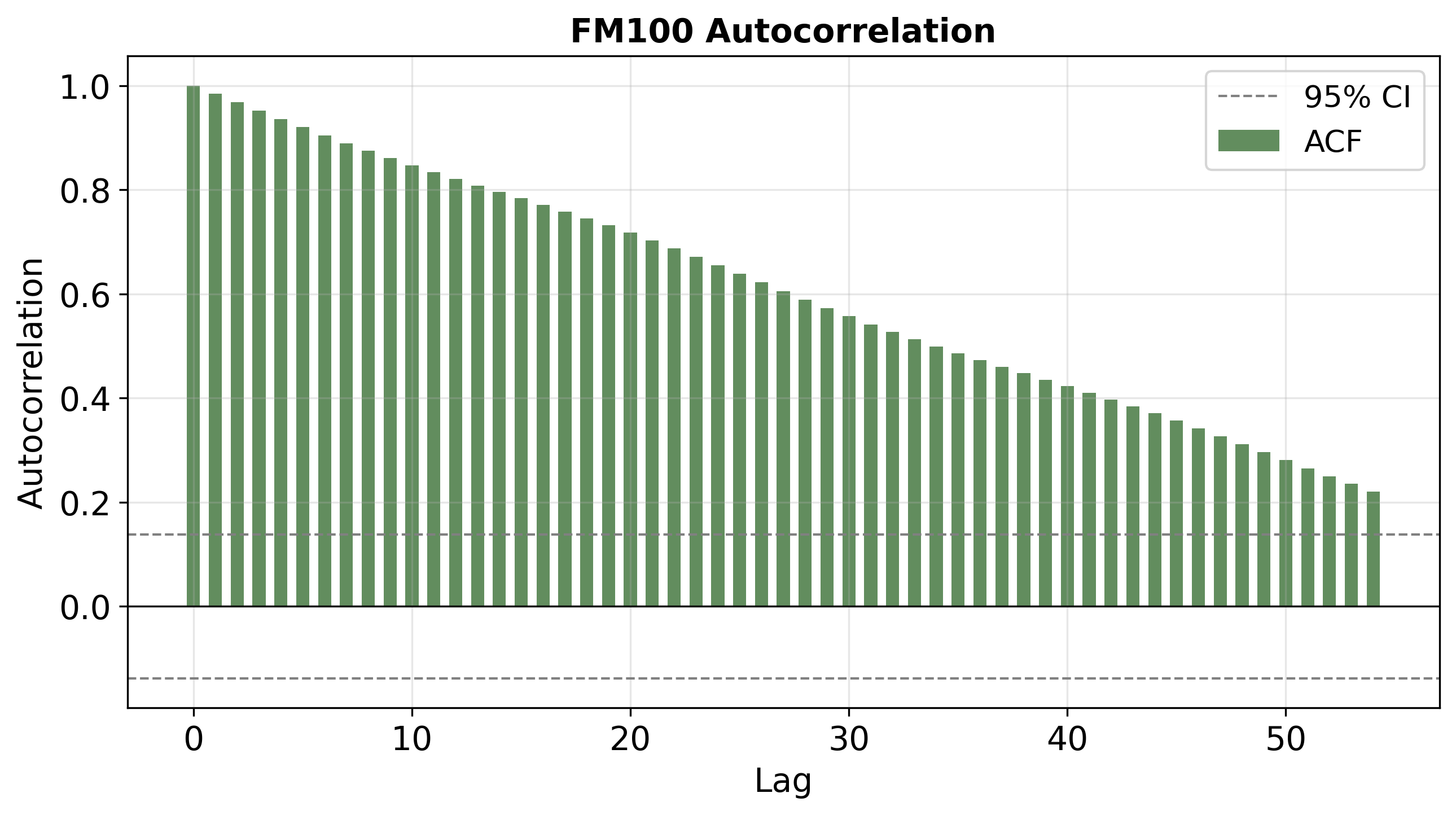}
\end{minipage}
\begin{minipage}{0.48\textwidth}
  \includegraphics[width=\linewidth,keepaspectratio]{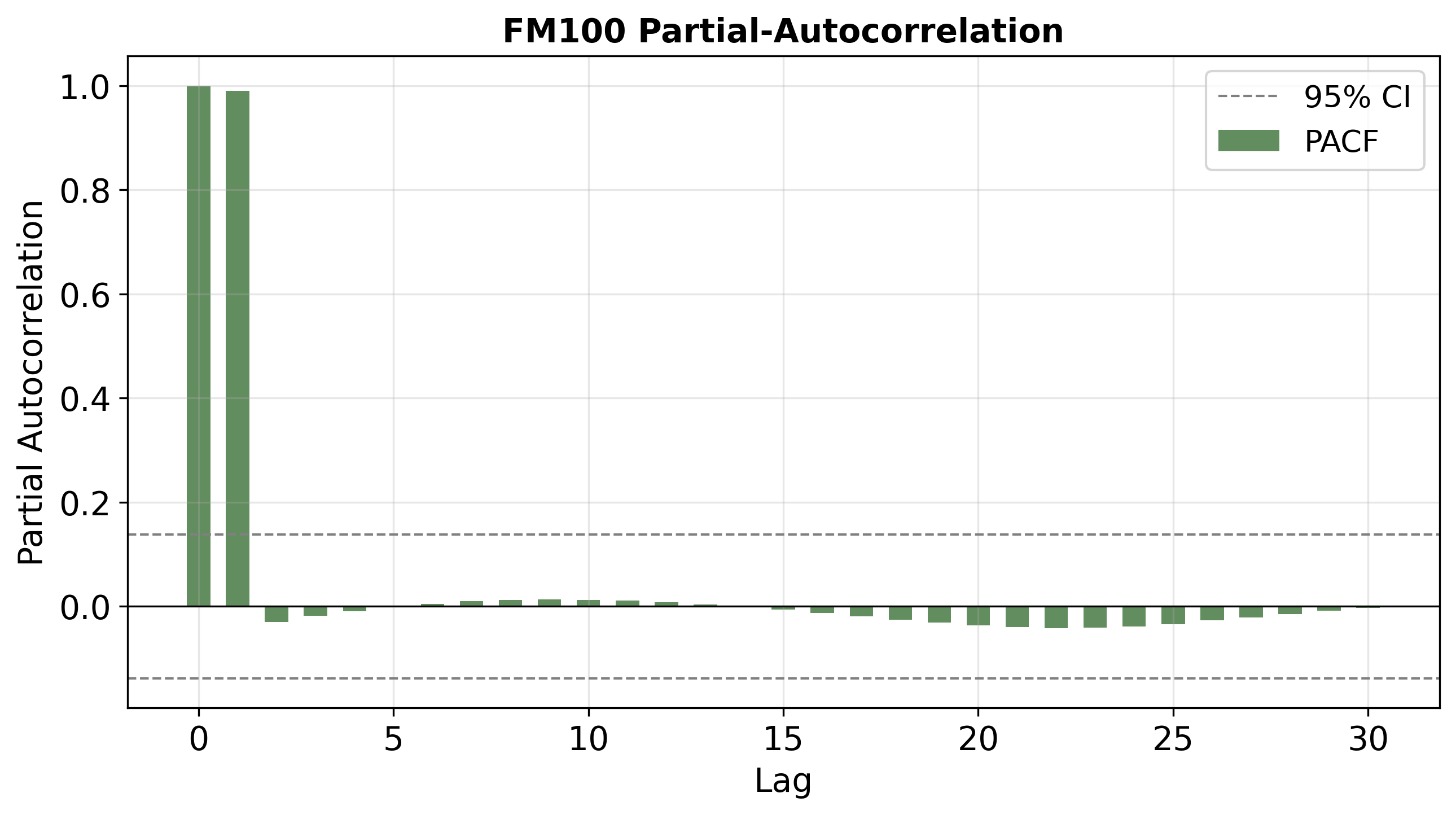}
\end{minipage}

\begin{minipage}{0.48\textwidth}
  \includegraphics[width=\linewidth,keepaspectratio]{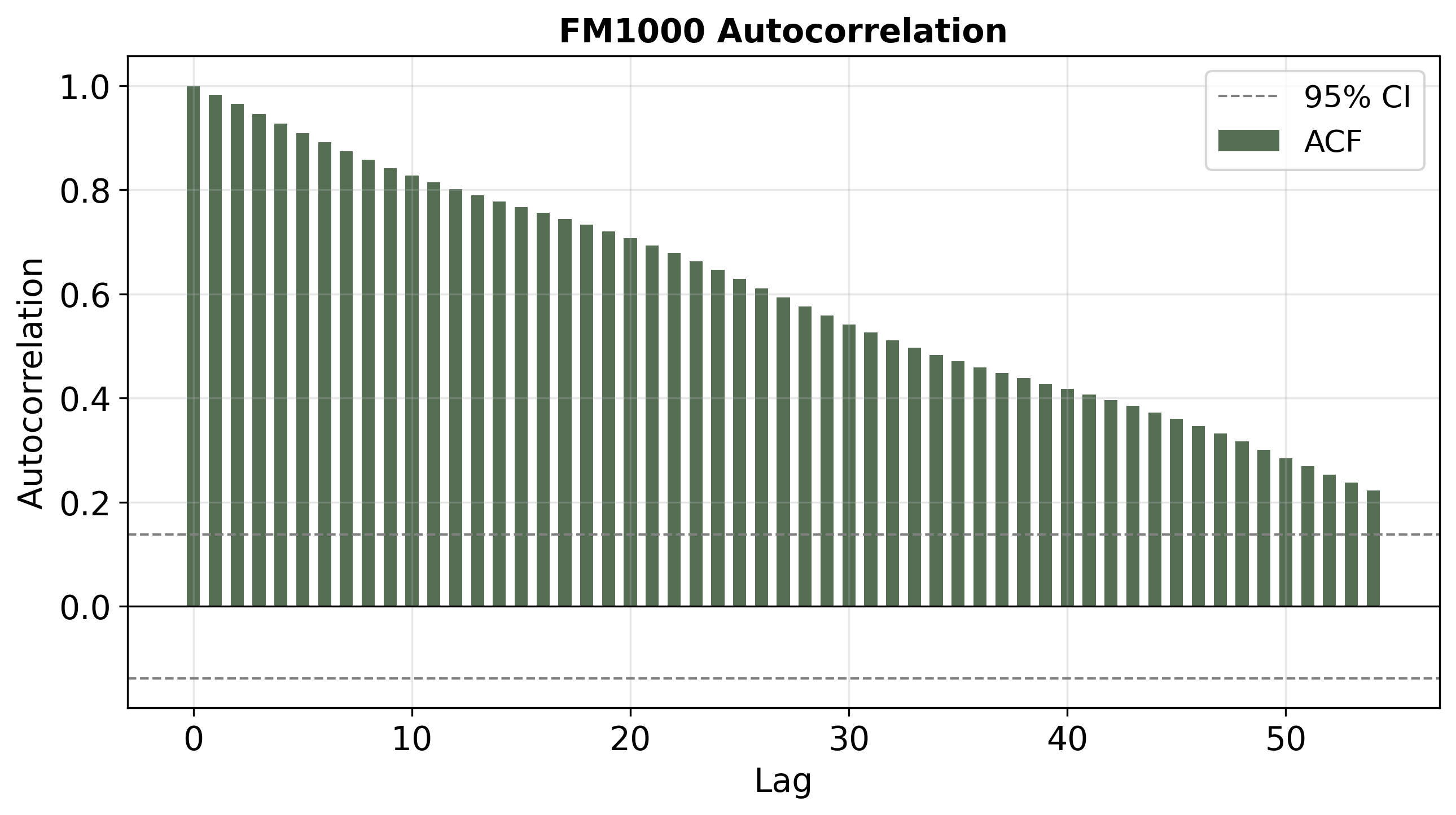}
\end{minipage}
\begin{minipage}{0.48\textwidth}
  \includegraphics[width=\linewidth,keepaspectratio]{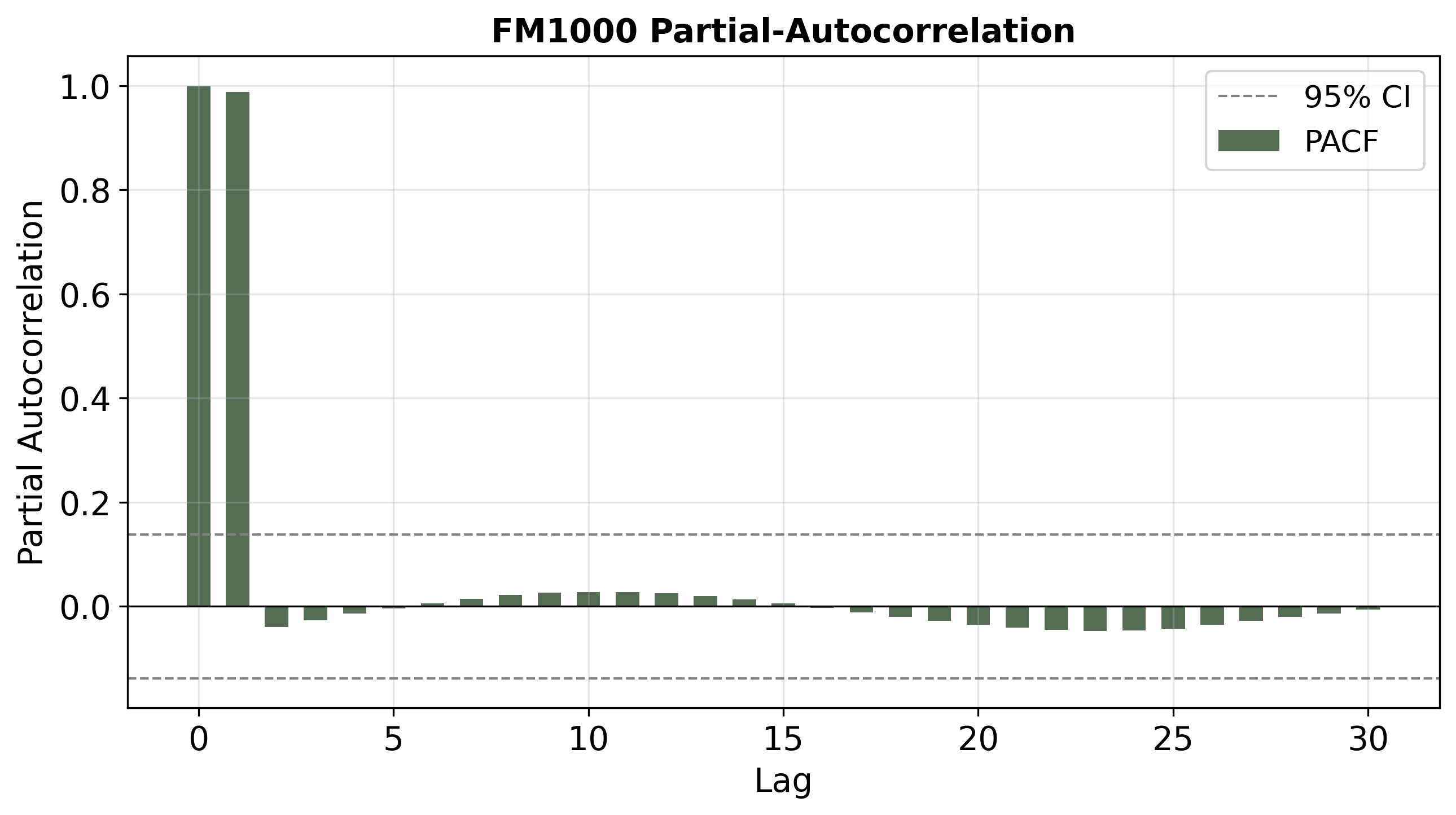}
\end{minipage}

\caption[ACF and PACF of FM1, FM100, and FM1000]{Autocorrelation (left column) and partial autocorrelation (right column) functions for FM1 (top row), FM100 (middle row), and FM1000 (bottom row) from predictions on the test set. The plots represent the ACF/PACF for the single models that had the median RMSE on the test set. The ACF for FM1 decays relatively quickly and has a visible diurnal periodicity, while FM100 and FM1000 show much slower decay consistent with stronger persistence in the system. FM100 and FM1000 show a dominant lag-1 effect consistent with strong persistence, whereas FM1 shows additional short-lag structure and weaker persistence beyond the first lag.}
\label{fig:acf_pacf_all}
\end{figure}

Next, we analyze the autocorrelation function (ACF) and the partial autocorrelation function (PACF) of the Time-Warping predictions in the Oklahoma test set \citep[Section 2.1]{Box-2016-TSA}. The ACF and PACF provide a statistical way to analyze the temporal structure of the predictions. Here, they are used to assess whether the time-warping method changes the learned dynamics in the intended way.
We used a single representative model for each fuel class, since averaging across realizations can smooth the autocorrelation structure.
The plotted values are the hourly RNN predictions rather than the observations, which are too sparse in time to analyze short-lag autocorrelation.
Figure~\ref{fig:acf_pacf_all} shows the ACF and PACF plots for FM1, FM100, and FM1000.

The ACF for the FM1 predictions shows a clear diurnal pattern, with negative correlation near 12-hour lags and positive correlation near 24-hour lags. By contrast, the ACFs for FM100 and FM1000 are very similar and resemble the gradual decay expected for an AR(1) process \citep[Sect.~3.2.3]{Box-2016-TSA}, indicating stronger persistence. The FM1 ACF decays much more quickly, which suggests weaker dependence on the previous state and greater influence from diurnal variability in the weather. Because there is no physical reason to expect FMC itself to become negatively autocorrelated in the absence of changing environmental forcing, the negative autocorrelation at shorter lags in FM1 is most plausibly explained by weather-driven variability.

The PACF for all three predictions shows a strong positive correlation at lag~1, consistent with the physical persistence of FMC. For FM1, the PACF is also negative at lags 2 to 4, which likely reflects short-term weather forcing from wetting and drying events. By contrast, the PACFs for FM100 and FM1000 drop below the significance threshold after lag~1, which is consistent with the PACF expected for an AR(1) process. Overall, the PACF results support a stronger role for external forcing in FM1 and greater persistence in FM100 and FM1000.

The ACF and PACF results indicate that FM1 is more strongly influenced by weather forcing, whereas FM100 and FM1000 are more strongly governed by persistence.

\section{Discussion, Conclusions, and Future Research}\label{sec:conclusions}

In this paper, we introduce a novel time-warping method for transfer learning that aims to modify the learned temporal dynamics of an RNN with LSTM recurrent layers. Using data from the influential Oklahoma field study, we found that an RNN pretrained with NWP inputs and FM10 sensor observations remained accurate when applied with ground-based weather inputs and gravimetric measurements. We compared the time-warping method with established transfer learning baselines and found that it produced consistent accuracy for FM1, FM100, and FM1000 despite fine-tuning only a small fraction of the parameters updated by the other methods.

The zero-shot FM10 results help establish that the pretrained RNN transfers knowledge reliably, with accuracy consistent with the variability of the source model across realizations and performance comparable to that of the Nelson model. The RNN had never been exposed to data from the Oklahoma field study, but for FM10 less than or equal to $30\%$ its accuracy was comparable to that of the state-of-the-art Nelson model. Although the zero-shot model performed poorly for very wet fuels, likely due to systematic biases in FM10 sensors, this result still suggests that the pretrained RNN learned weights that generalize well across locations, weather inputs, and observation methods. The Nelson model was specifically calibrated to the Oklahoma field study, whereas the RNN was applied there without fine-tuning after pretraining on different data sources. 

For FM1, the main result is that the Time-Warping method remained competitive with the standard transfer learning baselines, and there is reason to suspect that it is less prone to overfitting the sparse target data. For FM1, FM100, and FM1000, the time-warping method fit only two parameters, which shifted 128 out of more than 21,000 trainable parameters in the RNN. This suggests that much of the useful structure learned from FM10 can be retained while adapting only a small part of the model to the target fuel class. Because the observed data are sparse, this limited adjustment may be less prone to overfitting than methods that fine-tune many more parameters on a single location and time period. The time-warping parameters were optimized using data from a single location and year, so some overfitting remains possible. However, the adjustment was small relative to the full model complexity, and these parameters were chosen based on theoretical considerations rather than post-hoc tuning after seeing the target data. The ACF and PACF analyses further suggest that the method changes the temporal behavior in the intended direction, with FM1 showing stronger weather forcing and FM100/FM1000 showing greater persistence.

The other transfer learning methods fine-tuned many more parameters on the sparse target dataset, so overfitting in both space and time remains a greater concern. In the time-warped RNN, the input connections for longitude, latitude, and elevation remained unchanged along with most of the other parameters. Although modifying the forget and input gate biases could in principle reduce spatial interpolation accuracy, we hypothesize that the time-warping method may generalize more robustly to new locations and times than methods that fit many parameters to data from a single location.

The time-warping method therefore contributes to the literature in several ways. First, this paper frames time-warping as a form of transfer learning in which the input space is shared across fuel classes, but the prediction task differs by time scale. This provides a useful way to transfer the learned temporal structure between related prediction problems with different characteristic response times. Second, it shows that transfer learning is useful for fuel moisture prediction between fuel classes with very different data availability. To our knowledge, this is also the first use of a recurrent ML method for FM100 and FM1000 prediction. Third, the specific time-warping method developed here has the potential to support operational prediction of sparsely observed fuel classes.

An important direction for future research is broader evaluation across locations and times. The FEMS field samples were not used in this study because their temporal resolution is much lower than that of the Oklahoma field study, but they include observations from across CONUS and could support a broader comparison with the Nelson model than is possible with the relatively small test dataset used here. This would be an important test of generalization because the RNN was developed using FM10 sensor observations and NWP inputs rather than local gravimetric observations and station-based weather data, making the evaluation on the Oklahoma field study a relatively demanding transfer setting.

Additional future work should examine whether alternative loss functions can better capture the large FMC changes associated with infrequent rain events, and whether low-FMC conditions after recent rain should be evaluated separately from low-FMC conditions without recent rain, including comparison of NWP forcing with local station forcing during post-rain recovery. Computational cost should also be assessed, since ML methods may in some cases improve prediction speed.

\acknowledgments
The research presented in this paper was partially funded by NASA grants 80NSSC22K1717, 80NSSC23K1118, 80NSSC22K1405, 80NSSC23K1344, and 80NSSC25K7276. 

Computing resources were provided by the University of Colorado Denver supercomputing cluster Alderaan, partially funded by NSF Campus Cyberinfrastructure grant OAC-2019089.

A portion of this work used code generously provided by Brian Blaylock's Python packages Herbie (Version 2024.8.0) (\url{https://github.com/blaylockbk/Herbie}) and SynopticPy (\url{https://github.com/blaylockbk/SynopticPy}). 

Large Language Models, specifically ChatGPT-5, were utilized for support with coding, literature search, and writing revisions.

%
%
\datastatement
The code associated with this study is publicly available at \url{https://github.com/jh-206/fmc_transfer}. The source-task data were derived from public-domain resources, including the Amazon AWS archive of historical HRRR forecasts (\url{https://registry.opendata.aws/noaa-hrrr-pds/}) and historical RAWS observations from Synoptic, accessed through SynopticPy (\url{https://github.com/blaylockbk/SynopticPy}). The transfer-learning repository also contains the Oklahoma field study dataset used in this paper, together with the code and supporting files needed to reproduce the analysis. The Oklahoma Mesonet weather data associated with that study can be shared with proper citation. Additional processed datasets are available from the authors upon reasonable request, subject to university computing and storage constraints.



\appendix





%


\appendixtitle{Data and Code}

The code associated with this project was developed in Python 3.11.14, primarily using TensorFlow 2.20.0 and Keras 3.12.0. Two publicly available GitHub repositories form the code base of the project. 

The repository \url{https://github.com/openwfm/ml_fmda} contains the materials needed to reproduce the dataset used in the source learning task and the associated results reported primarily in \citet{Hirschi-2026-RNN}. The data were derived from public resources, including the Amazon AWS archive of HRRR forecasts (\url{https://registry.opendata.aws/noaa-hrrr-pds/}) and  RAWS observations from Synoptic, accessed through SynopticPy (\url{https://github.com/blaylockbk/SynopticPy}). These data were collected, formatted, and stored using the University of Colorado Denver supercomputing cluster Alderaan. The exact processed datasets are available from the authors upon reasonable request, subject to university computing and storage constraints.

The GitHub repository \url{https://github.com/jh-206/fmc_transfer} contains the code and supporting files needed to reproduce the transfer-learning analysis, including the Oklahoma field-study dataset used in this paper. The repository uses 100 RNN replications that are archived separately online~\citep{Hirschi-2026-RRF}.

The Oklahoma field study data include hourly weather observations from a ground-based station in the Oklahoma Mesonet~\citep{OKMeso-2026-MSI}. These data were purchased from Mesonet and can be shared with proper citation. The wetting and drying equilibria were calculated from the Mesonet air temperature and relative humidity. Because the Mesonet observations were reported directly on the hour while the FMC measurements sometimes included minutes within the hour, we aligned the times for the RNN predictions and accuracy calculations as follows.

Times were converted from Central Time to UTC so that the Oklahoma field study data were consistent with the RAWS and HRRR data used in the source learning task. This was also necessary because hour of day and day of year were used as derived predictors in the RNN. For the training set, weather variables from the nearest hour were assigned to each FMC observation, introducing a possible temporal misalignment of up to half an hour. For the test set, the hourly RNN predictions were linearly interpolated to the exact times of the FMC observations for the accuracy calculations.

\bibliographystyle{ametsocV6}
\bibliography{references}

\end{document}